\documentclass{aa}
\usepackage{txfonts}
\usepackage{graphicx}
\usepackage{natbib}
\usepackage{color}

\begin{document}

\title{Spatially extended emission around the Cepheid RS Puppis in near-infrared hydrogen lines}
\titlerunning{Spatially extended emission around RS Puppis}

\subtitle{Adaptive optics imaging with VLT/NACO\thanks{Based on observations made with ESO telescopes at Paranal observatory under program ID 382.D-0065(A)}}

\author{ A.~Gallenne\inst{1,2} \and  
  A.~M\'erand\inst{1} \and 
  P.~Kervella\inst{2} \and
  J. H. V. Girard\inst{1} }
  
\authorrunning{A. Gallenne et al.}

\institute{ European Southern Observatory, Alonso de C\'ordova 3107,
  Casilla 19001, Santiago 19, Chile
  \and LESIA, Observatoire de Paris, CNRS UMR 8109, UPMC, Universit\'e
  Paris Diderot, 5 Place Jules Janssen, F-92195 Meudon, France}
  
  \offprints{A. Gallenne} \mail{alexandre.gallenne@obspm.fr}
       
\abstract
	{It has been recently discovered that Cepheids harbor circumstellar envelopes (CSEs).
	RS~Pup is the Cepheid that presents the most prominent circumstellar envelope known, the origin of which is not yet understood.} 
   {Our purpose is to estimate the flux contribution of the CSE around RS~Pup at the one arcsecond scale ($\sim 2000\,\mathrm{AU}$) and to investigate its geometry, especially regarding asymmetries, to constrain its physical properties.}
   {We obtained near-infrared images in two narrow band filters centered on $\lambda = 1.644$ and $2.180\,\mu$m (NB\_1.64 and IB\_2.18, respectively) that comprise two recombination lines of hydrogen: the 12--4 and 7--4 (Brackett $\gamma$) transitions, respectively. We used NACO's cube mode observations in order to improve the angular resolution with the shift-and-add technique, and to qualitatively study the symmetry of the spatially extended emission from the CSE with a statistical study of the speckle noise.} 
   {We probably detect at a 2$\sigma$ level an extended emission with a relative flux (compared with the star in the same filter) of $38\,\pm\,17\%$ in the NB\_1.64 filter and $24\,\pm\,11\%$ in the IB\_2.18 filter. This emission is centered on RS~Pup and does not present any detectable asymmetry. We attribute the detected emission to the likely presence of an hydrogen envelope surrounding the star.}
   {}


\maketitle


\section{Introduction}

For almost a century, Cepheid stars have been used as distance indicators thanks to the well-known period-luminosity relation (P-L), also called Leavitt's law. A good calibration of this relation is necessary to obtain an unbiased estimate of the distance. 
After the detection of circumstellar envelopes (CSEs) around many Cepheids in the near- and mid-infrared \citep{Kervella-2006-03, Merand-2006-07, Merand-2007-08, Kervella-2009-05}, it was shown that these CSEs could bias the angular diameter determination using near-infrared interferometry \citep{Merand-2006-07}. 
A circumstellar envelope could have an impact on the distance estimate if we do not take into account its flux contribution. It could lead to a bias both in interferometric and photometric measurements and therefore influence the accuracy of the estimated distance. Because Cepheids are used as standard candles to estimate distances in the universe, it is important to understand the properties of the CSEs around this kind of stars.

We report new observations with VLT/NACO in the IB\_2.18 and NB\_1.64 filters using a fast imaging mode in order to characterize the extended emission around \object{RS~Puppis} (HD 68860) at sub-arcsecond angular resolution	($\approx$0.01\,pc at RS~Pup distance). Data and reduction methods are described in Sect. \ref{observations-and-data-reduction}. In Sects. \ref{shiftandadd} and \ref{morphological-analysis} we present two different data processing analyses: the shift-and-add method, and a method based on a statistical analysis of the speckle noise to enable the detection of the CSE in the speckle cloud surrounding the star's image.

This paper is focused on RS~Pup's envelope, which has been known for many years \citep{Westerlund-1961-02,Havlen-1972-01} for it is very extended: it has a spatial extension of around 2\arcmin\ in V band, which corresponds to about one parsec at RS~Pup's distance \citep[about 2 kiloparsecs;][]{Kervella-2008-03}. Its origin is not precisely known; it is especially unclear whether it is made of material ejected by the star or of a material that predates the formation of the star. Its actual shape has been widely discussed: discrete nebular knots have been invoked by \citet{Kervella-2008-03} to estimate the distance using light echoes, whereas \citet{Feast-2008-06} argued that a disk-like CSE will bias this kind of estimate. Alternately, a bipolar nebula has been considered by \citet{Bond-2009-02}, making a ratio between two images of different epochs. We were able to spatially resolve the CSE close to the star. The morphology of the CSE is discussed in Sect. \ref{shape-of-the-cse}.



\section{Observations and data reduction}
\label{observations-and-data-reduction}

RS~Pup was observed with the NACO instrument installed at the Nasmyth B focus of UT4 of ESO VLT. NACO is an adaptive optics system \citep[NAOS, ][]{Rousset-2003-02} and a high-resolution near IR camera \citep[CONICA, ][]{Lenzen-2003-03}, working as imager or as spectrograph in the range 1--5\,$\mathrm{\mu m}$. We used the S13 camera for our RS~Pup observations \citep[FoV of 13.5\arcsec$\times$13.5\arcsec\ and $13.26 \pm 0.03$ mas/pixel,][]{Masciadri-2003-11} with the narrow band filter NB\_1.64 ($\lambda_0$ = 1.644\,$\mathrm{\mu m}$ with a width $\Delta \lambda$ = 0.018\,$\mathrm{\mu m}$) and the intermediate band filter IB\_2.18 ($\lambda_0$ = 2.180\,$\mathrm{\mu m}$ with a width $\Delta \lambda$ = 0.060\,$\mathrm{\mu m}$). We chose the cube mode in order to apply the shift-and-add technique. 
We chose windows with 512$\times$514 pixels and 256$\times$258 pixels respectively for the narrow and the intermediate band filters. Each cube (10 for RS~Pup and 4 for the reference star) contains 460 images at 1.64\,$\mu$m and 2~000 at 2.18\,$\mu$m. Each frame has the minimum integration time and readout type for this mode, i.e 109\,ms and 39\,ms respectively for the 1.64 and 2.18\,$\mu$m cubes. With these DITs and filters we are in a quasi-monochromatic regime with a very good speckle contrast \citep{Roddier-1981-}. The large number of exposures allows us to select the frames with the brightest image quality. 

Data were obtained on 2009 January 7, with the atmospheric conditions presented in Table~\ref{atmospheric_conditions}. The PSF (point spread function) calibrator stars were observed just before or immediately after the scientific targets with the same instrumental configurations. Each raw image was processed in a standard way using the Yorick\footnote{http://www.maumae.net/yorick/doc/index.php} language: bias subtraction, flat-field and bad pixel corrections. The negligible sky background was not subtracted.

To compare the AO system behavior and artefacts on RS~Pup, we used the star Achernar and its PSF reference star observed the same night. These data were also acquired in cube mode with the same camera in the NB\_1.64 and NB\_2.17 ($\lambda_0$ = 2.166\,$\mathrm{\mu m}$ with a width $\Delta \lambda$ = 0.023\,$\mathrm{\mu m}$) filters. Only one cube was acquired per filter, containing 20~000 frames for Achernar and 8~000 for the calibrator. The window detector size is 64$\times$66 pixels in both filters with an integration time of 7.2\,ms. The slightly different second filter and the shorter exposure time compared with RS~Pup is not critical for our analysis because we are interested in flux ratios. The smaller window size cuts the AO PSF wings, but we will show in Sect.~\ref{shiftandadd} that this does not affect our conclusion significantly.

\begin{table*}[ht]
\centering
\begin{tabular}{cccccccccccc} 
\hline
\hline
Date				& UTC (start-end)			& 	$\phi$	& Filter 			& Name 						& $\mathrm{\overline{seeing}_{DIMM}}$ & $\mathrm{\overline{seeing_\lambda}}$ 	& $\mathrm{AM}$	&	Nbr & $\overline{r_0}$ (cm) & $\overline{t_0}$ (ms) \\
\hline
2009-01-07	&	00:30:41-00:31:56	&		-		& NB\_2.17 	& HD~9362               	& 0.72\arcsec\ 							&0.55\arcsec\           				& 1.15 		& 1 	& 9.2						& 3.1						\\
2009-01-07	& 	00:35:33-00:36:47	&		-		& NB\_1.64 	& \object{HD~9362} 		& 0.77\arcsec\ 							& 0.62\arcsec\           				& 1.15 		& 1 	& 9.0						& 2.9						\\ 
2009-01-07	&	00:51:16-00:54:16	&		-		& NB\_2.17 	& Achernar                	& 0.84\arcsec\ 							& 0.64\arcsec\           				& 1.25		& 1 	& 9.1						& 2.3						\\
2009-01-07	&	01:01:30-01:04:28	&		-		& NB\_1.64 	& \object{Achernar}  		& 0.83\arcsec\ 							& 0.63\arcsec\           				& 1.27 		& 1 	& 9.1						& 2.3						\\ 
2009-01-07	&	04:47:17-05:00:46	&	0.035	& IB\_2.18   	& RS~Pup                   	& 0.71\,$\pm$\,0.05\arcsec\ 	& 0.54\,$\pm$\,0.04\arcsec\ 	& 1.03 		& 10	& 13.6\,$\pm$\,0.4	& 3.6\,$\pm$\,0.2	\\
2009-01-07	& 	05:01:16-05:13:17	&	0.035	& NB\_1.64 	& RS~Pup                      	& 0.68\,$\pm$\,0.04\arcsec\ 	& 0.54\,$\pm$\,0.03\arcsec\ 	& 1.03		& 10 & 13.0\,$\pm$\,0.1	& 3.4\,$\pm$\,0.1	\\	
2009-01-07	&	05:21:03-05:26:27	&		-		& IB\_2.18 	& HD~74417                 & 0.55\,$\pm$\,0.01\arcsec\ 	& 0.42\,$\pm$\,0.01\arcsec\ 	& 1.05	 	& 4 	& 19.3\,$\pm$\,0.8	& 5.1\,$\pm$\,0.4	\\
2009-01-07	&	05:26:57-05:31:43	&		-		& NB\_1.64 	& \object{HD~74417}  	& 	0.54\,$\pm$\,0.01\arcsec\ 	& 0.43\,$\pm$\,0.01\arcsec\ 	& 1.04	 	& 4 	& 20.3\,$\pm$\,0.1	& 5.1\,$\pm$\,0.1	\\ 
\hline
\end{tabular}
\caption{Log of the NACO observations. $\mathrm{\overline{seeing}_{DIMM}}$ are the seeing measurements in $V$ from the Differential Image Motion Monitor (DIMM) station, while $\mathrm{\overline{seeing}_\lambda}$ are interpolated to our wavelength. The uncertainties denote the standard deviations over the number of cubes. The start-end UTC indicated denotes the time when the first and the last cube was recorded. The phase $\phi$ of the Cepheid was computed using ($P,T_0$) from \citet{Kervella-2008-03}. AM denotes the airmass, while Nbr is the number of cubes recorded. $r_0$ and $t_0$ are the Fried parameter and the coherence time given by the AO real time calculator.}
\label{atmospheric_conditions}
\end{table*}

Once the pre-processing was over, we carried out a precentering as well as a sorting according to the maximum intensity of the central peak to reject the images for which the correction was the least efficient. On the one hand, we applied the shift-and-add technique to obtain the best angular resolution possible. On the other hand, we applied a statistical study aimed at extracting a possible diffuse (i.e. completely resolved) component inside of the speckle cloud of the central core (the star). These methods are explained in Sects.~\ref{shiftandadd} and \ref{morphological-analysis}.


\section{Shift-and-add process}
\label{shiftandadd}

In this section we use the shift-and-add method and then compare the RS~Pup images with another star observed on the same night for which we do not expect an envelope.

\subsection{RS~Pup}

Both stars, RS~Pup and HD~74417, are unresolved by the telescope: diffraction limits for an 8.2m UT are 55\,mas and 41\,mas in filters IB\_2.18 and NB\_1.64 respectively, while angular diameters are approximately one milli-arcsecond (1\,mas) for both stars.

The shift-and-add method was originally proposed by \cite{Bates-1980-03} as a technique for speckle imaging. Applied to NACO's cube mode, it enhances the Strehl ratio by selecting the frames that are the least altered by the atmospheric turbulence. It also reduces the halo contribution when used with adaptive optics by selecting the best AO-corrected frames. The efficiency of the cube mode vs. the standard long exposure mode is discussed in \citet{Kervella-2009-09}. \citet{Girard-2010-07} showed for the RS~Pup's reference star at 2.18\,$\mu$m that by using this method we have an absolute gain of 9\,\% in Strehl ratio.

Our processing steps are as follows: we select the 10\,\% best frames according to the brightest pixel (as a tracer of the Strehl ratio) in our 10 data cubes (4 for the calibrator). We then spatially resample them by a factor 4 using a cubic spline interpolation and co-align them using a Gaussian fitting on the central core, at a precision level of a few milliarcseconds \citep[this method is described in detail in][]{Kervella-2009-05}. Each cube is then averaged to obtain ten final average images (4 for the calibrator). We finally compute the mean of the 10 average images of RS~Pup (or the 4 for the calibrator).

\begin{figure*}[]
\centering
\includegraphics[width=6.6cm]{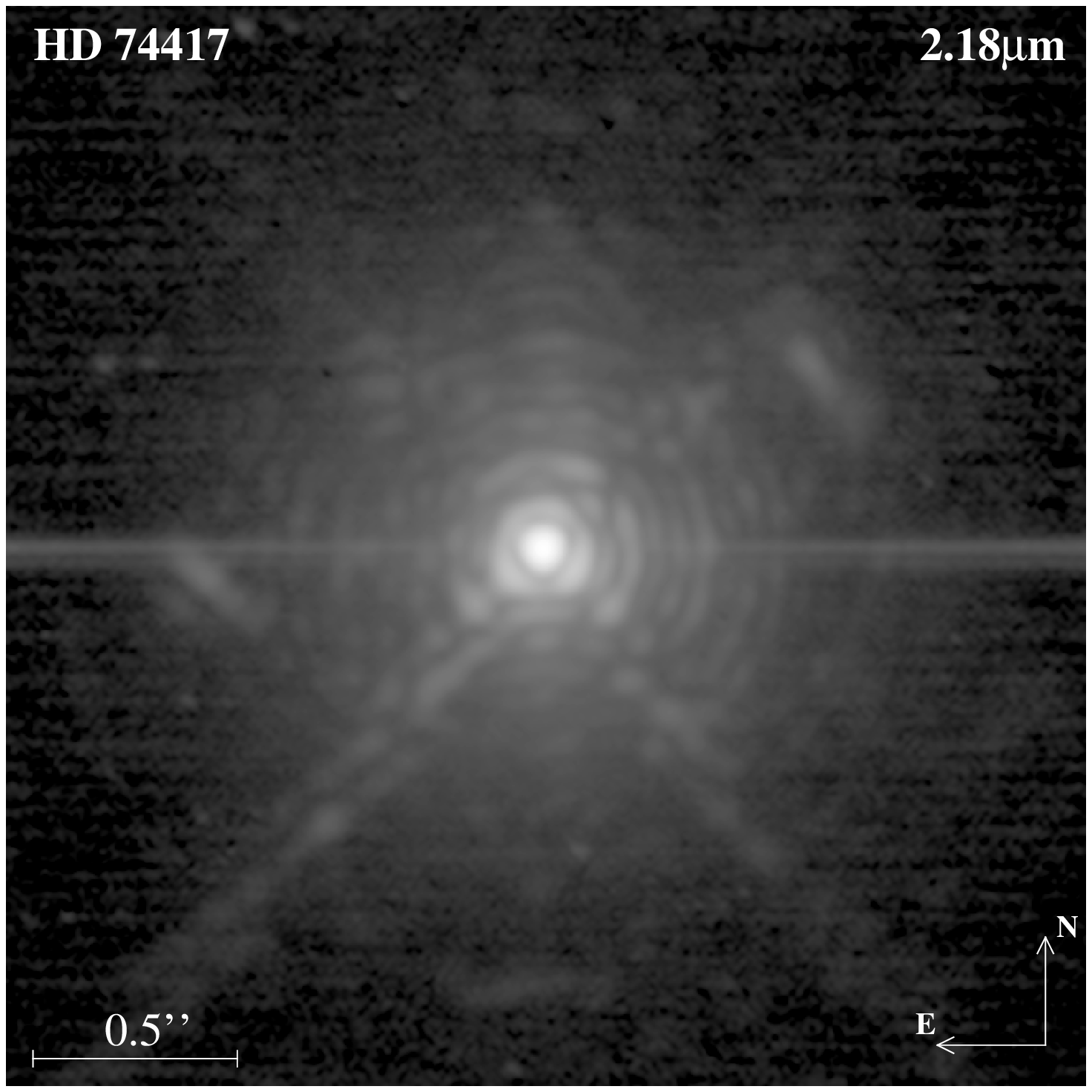}\hspace{.5cm}
\includegraphics[width=6.6cm]{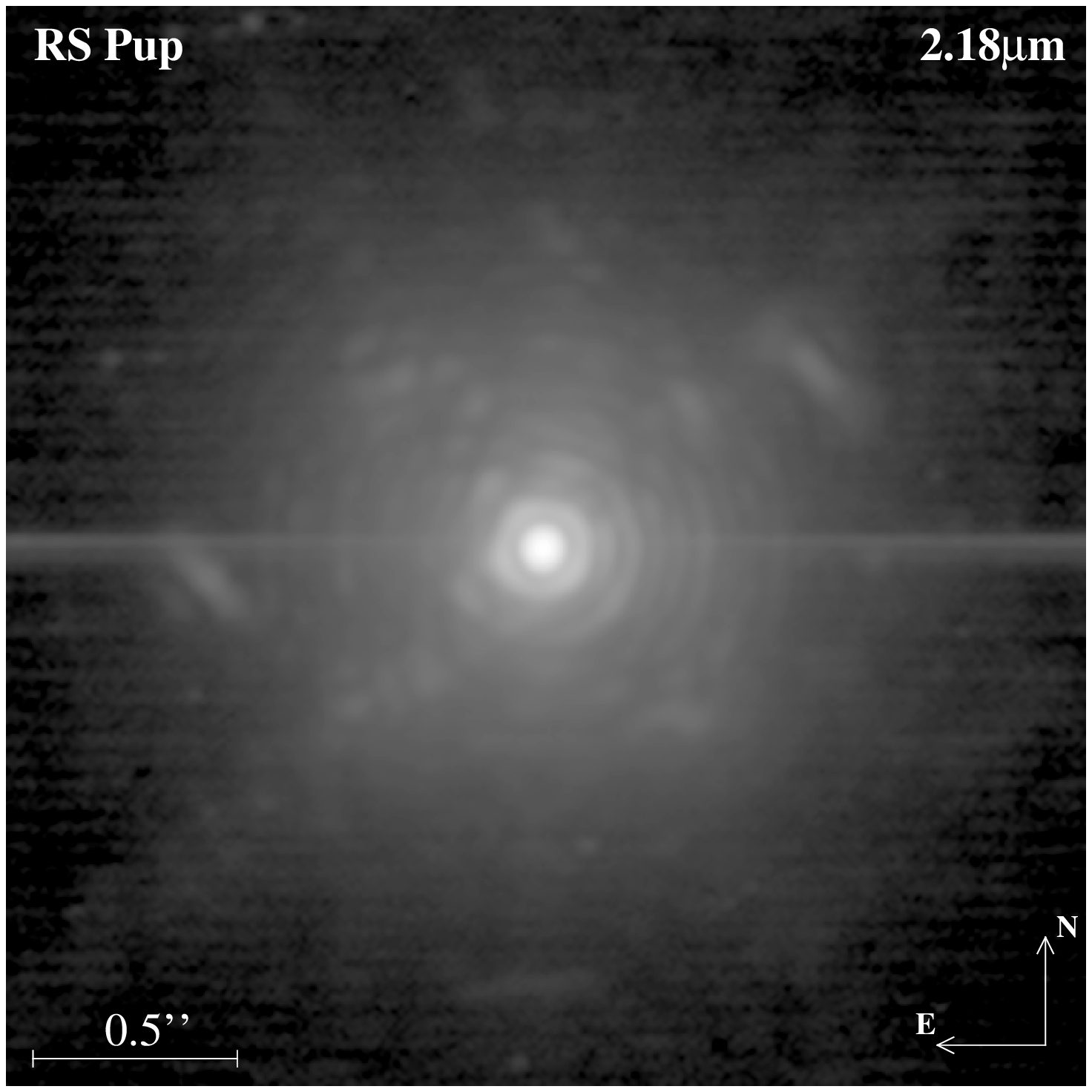}\vspace{.5cm}
\includegraphics[width=6.6cm]{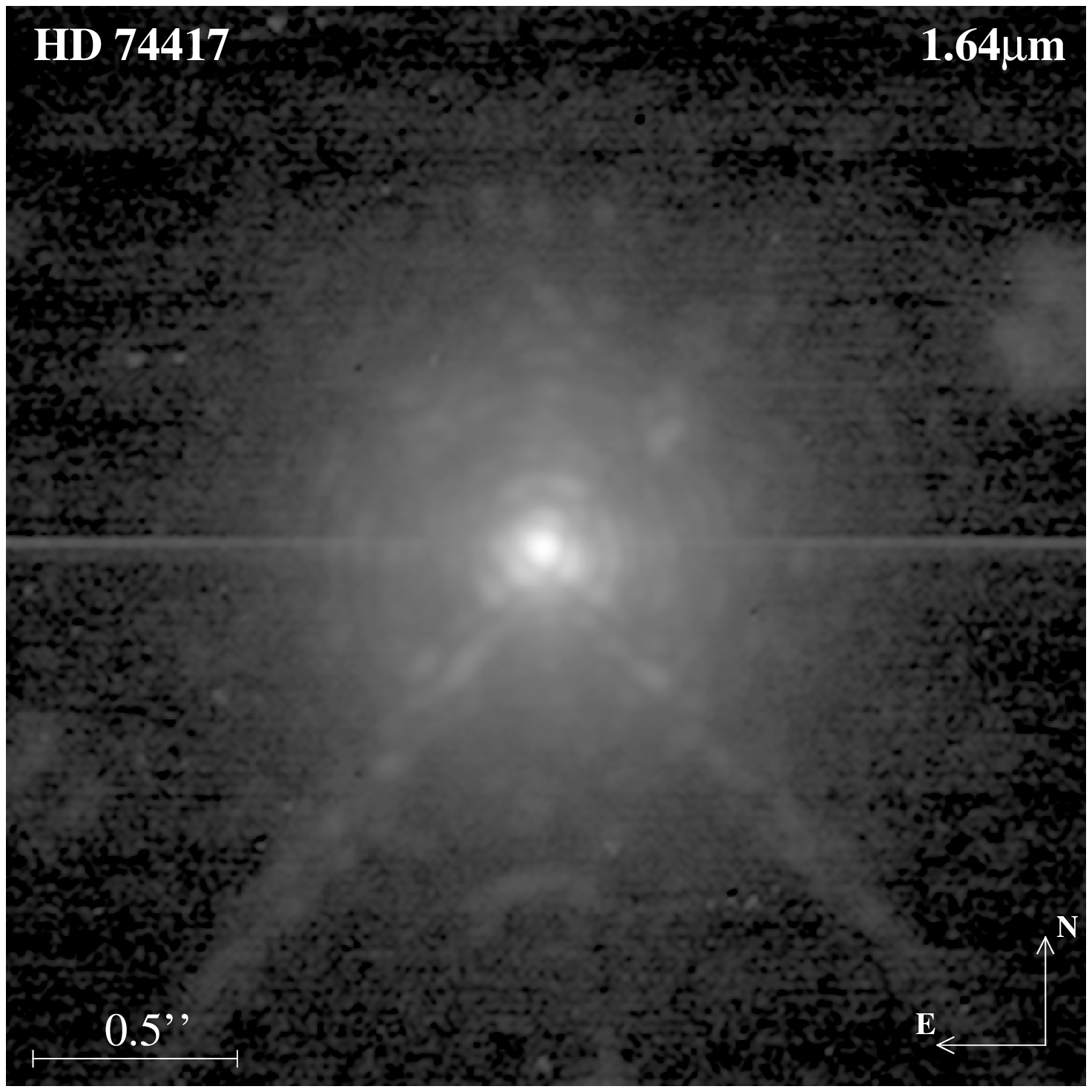}\hspace{.5cm}
\includegraphics[width=6.6cm]{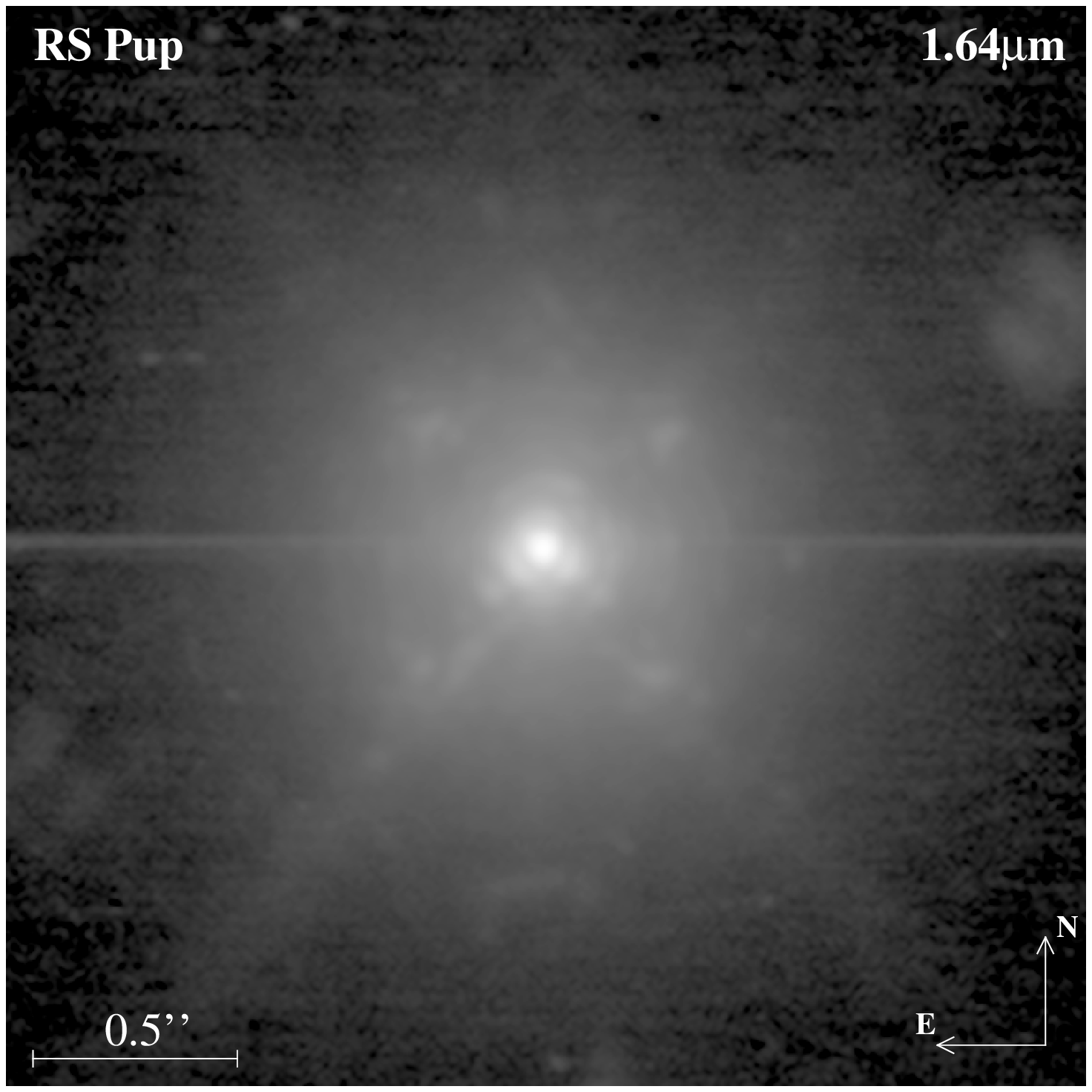}
\caption{Final frames from shift-and-add process. In the two upper panels the PSF calibrator star HD~74417 and RS~Pup are shown in the IB\_2.18 filter. The bottom panels show the same stars in the NB\_1.64 filter. The scale intensity is logarithmic.}
\label{images_shift_and_add}
\end{figure*}

The result of this method is shown in Fig.~\ref{images_shift_and_add} for the two filters. The two upper frames are  HD~74417 and RS~Pup as seen through the IB\_2.18 filter, while the two lower are in the NB\_1.64 filter. The spatial scale in all images is 3.3\,mas/pix and we chose a logarithmic intensity contrast with adjusted lower and higher cut-off to show details in the halo. On the 1.64\,$\mathrm{\mu m}$ image of RS~Pup compared with the HD~74417 image we can see that there is a resolved circumstellar emission up to $\sim$\,1\arcsec\ from the Cepheid. This feature is fainter in the 2.18\,$\mathrm{\mu m}$ filter. In both filters, because of this emission, diffraction rings are less visible around RS~Pup than around HD~74417.

For each final co-added frame (i.e the image shown in Fig.~\ref{images_shift_and_add}) we computed a ring median and normalized to unity. The resulting curves are presented in Fig.~\ref{radial_profil} for two filters. The error bars are the standard deviation of values inside rings and are plotted only for few radii for clarity. We notice a significant difference between the two curves, more pronounced in the NB\_1.64 filter.

\subsection{Achernar}

We carried out the same processing for Achernar and its reference star $\delta$~Phe (HD~9362). 
These two stars are unresolved by the UTs (angular diameters are approximately 2\,mas).
With this window size (64$\times$66 pixels) the field of view is four times smaller than in the RS~Pup's frames, and even if we can also see some Airy rings in all the images, the background is dominated by the AO halo.

\begin{figure*}[]
\centering
\includegraphics[width=6.6cm]{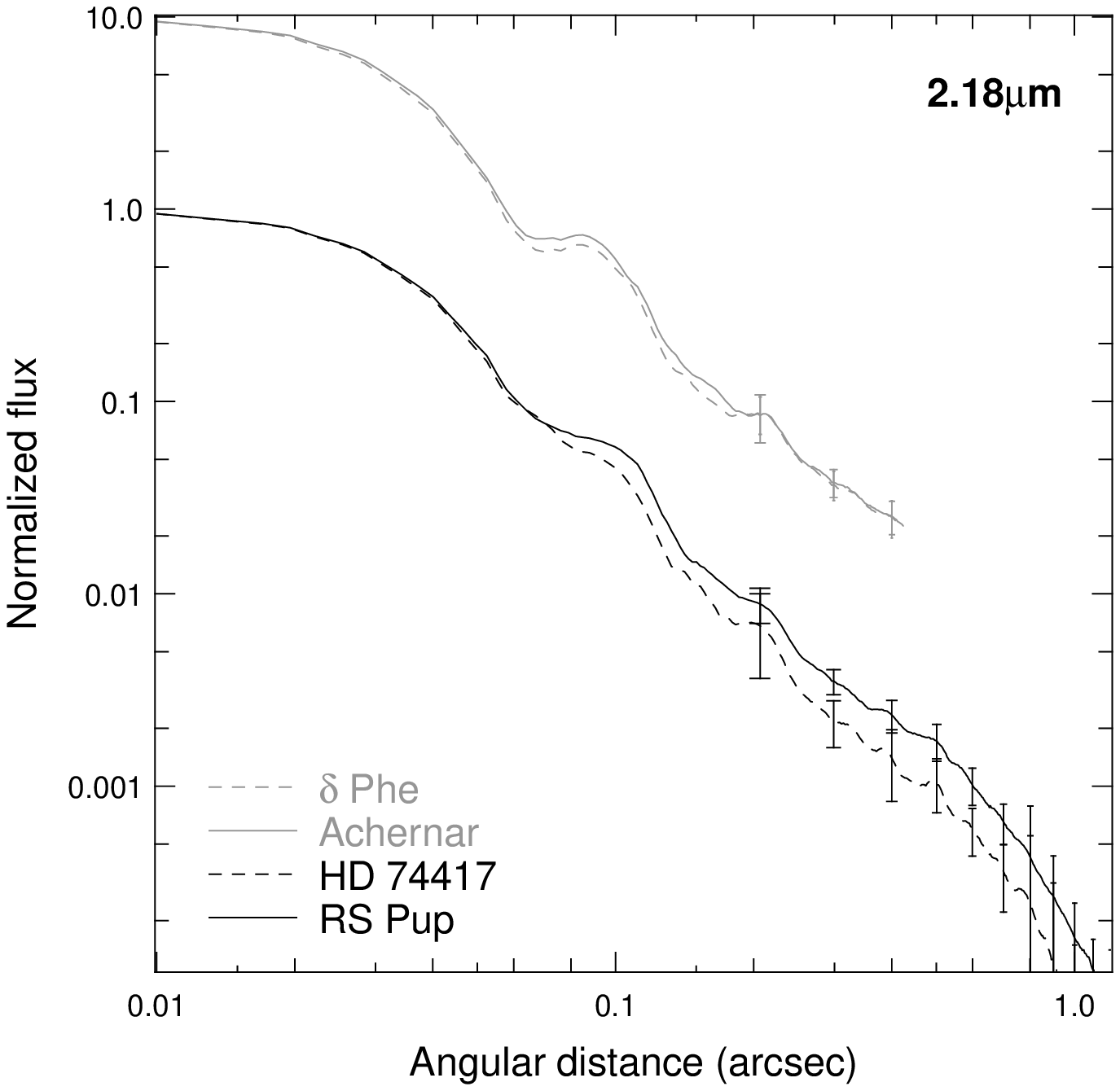}\hspace{.5cm}
\includegraphics[width=6.6cm]{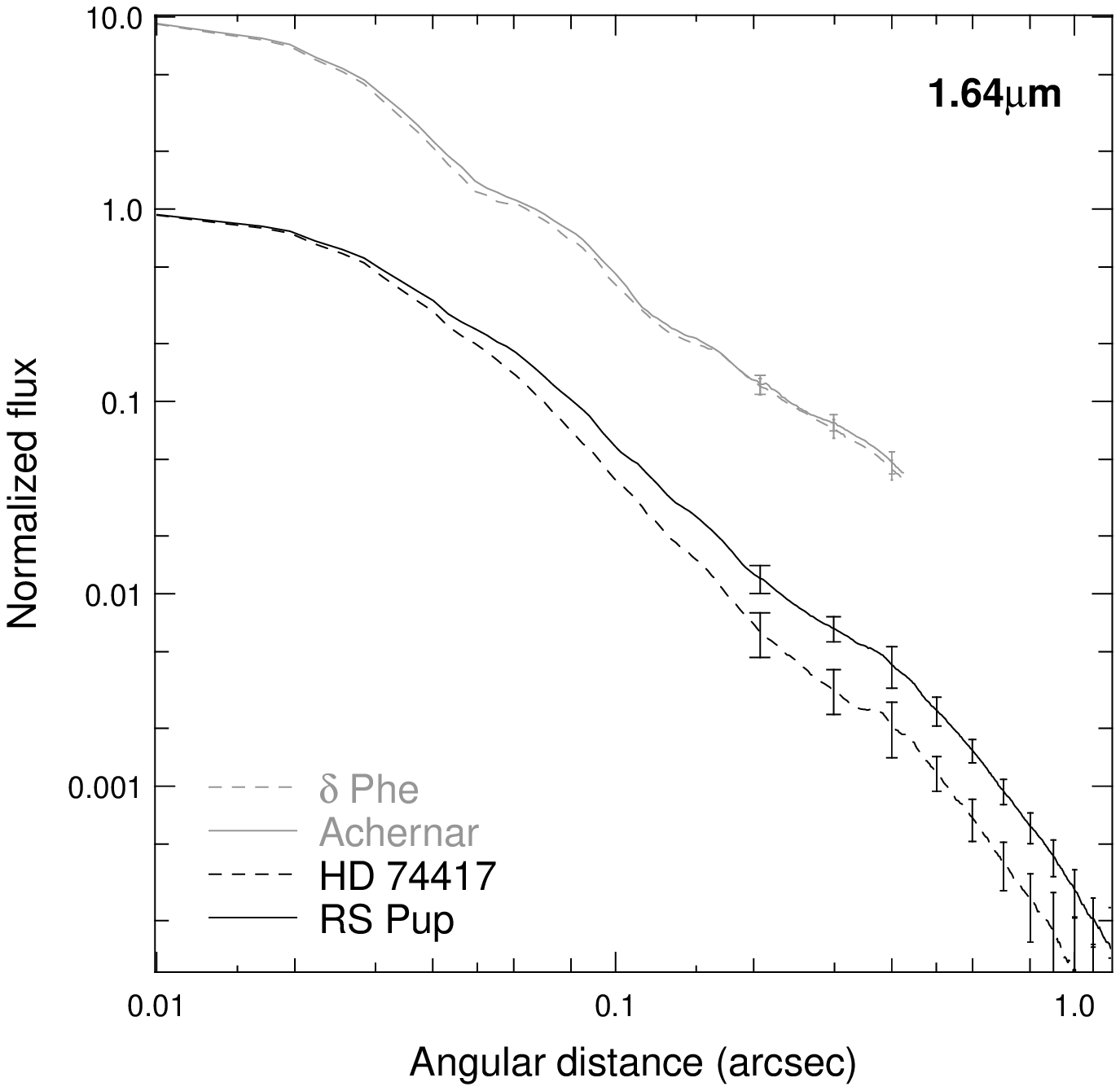}
\caption{On the left are presented the radial profiles at 2.18\,$\mu$m for RS~Pup, Achernar and their reference stars. On the right the same at 1.64\,$\mu$m. Radial profiles were computed using the median value on one-pixel thick circular rings. The gray color denotes Achernar and $\delta$ Phe profiles, shifted by a factor 10 for clarity. The scale is logarithmic for both axes.}
\label{radial_profil}
\end{figure*}

We plot in Fig.~\ref{radial_profil} the radial profiles of the stacked images, normalized to 10 (for clarity in the graph).

\subsection{Halo fitting}

We now investigate the presence of a circumstellar envelope by studying the radial profile of the stacked images we obtained for both RS~Pup and its PSF calibrator and compare them with the observations of Achernar. The main idea behind this study is to compare the variations of encircled energy (the ratio of the core's flux to the total flux) to the expected variations caused by atmospheric condition changes. If RS~Pup has indeed a circumstellar envelope, its encircled energy will be significantly lower than that of our reference stars.

We use an analytical function to fit the residual light out of the coherent core (i.e. the halo). We used a turbulence degraded profile from \citet{Roddier-1981-}

\begin{equation}
I_{\mathrm{halo}}(r)=f\ \frac{0.488}{\rho^2}\ \left[ 1+\frac{11}{6}\ \left( \frac{r}{\rho}\right) ^{2}\right]^{-\frac{11}{6}},
\label{I-halo}
\end{equation}
where the fitting parameters are the FWHM $\rho$ and $f$ proportional to the flux of the envelope. In this case, the parameter $\rho$ corresponds to the equivalent seeing disk for long exposure. NACO's resulting core, even for an unresolved source, is fairly complex and suffers from residual aberrations, this is why we decided to fit the halo at a large radial distance of the core. We will later define the core as the residual of the fit of the halo function to our data.

\begin{figure}[]
\centering
\includegraphics[width=4.45cm]{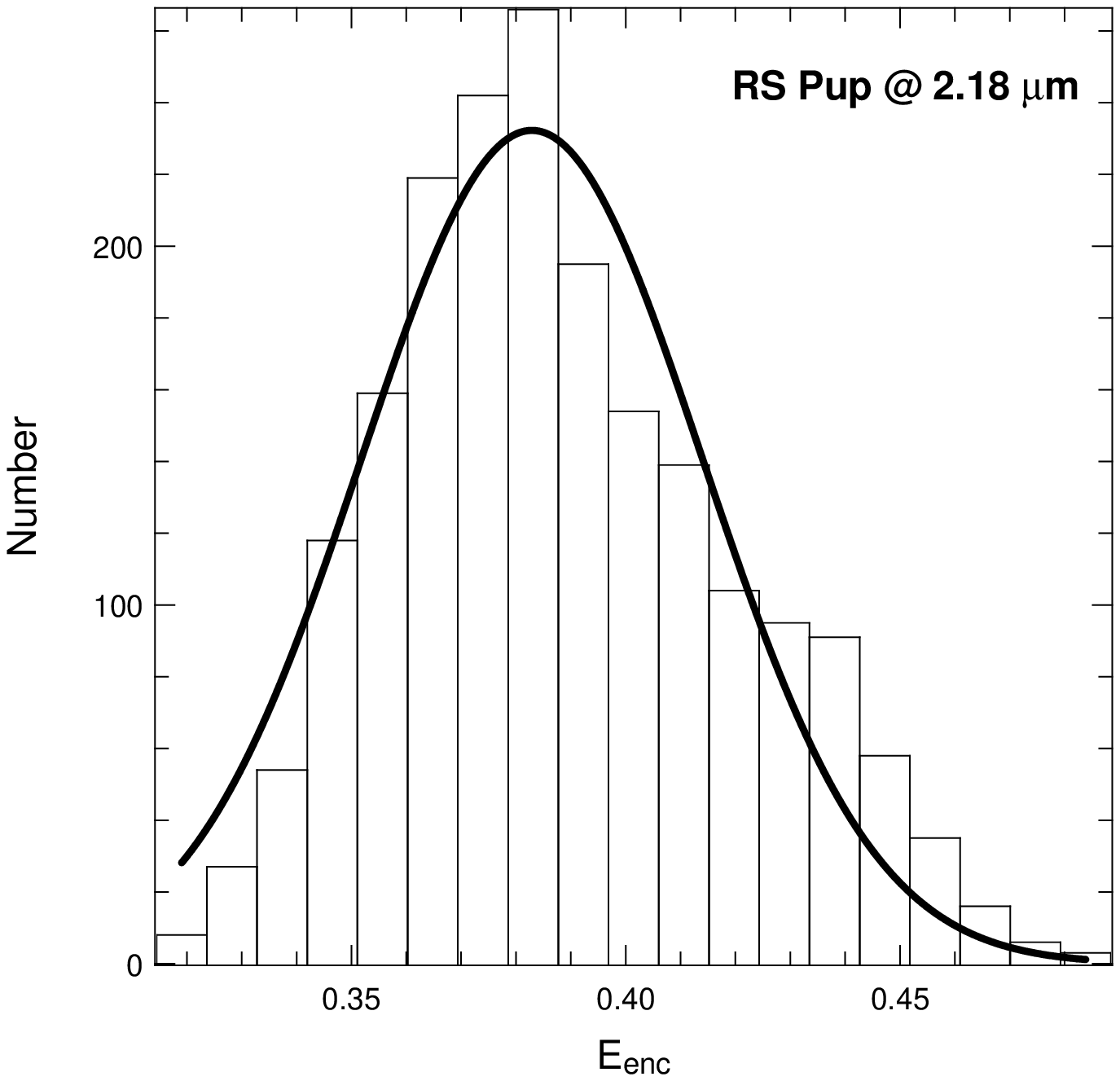}
\includegraphics[width=4.45cm]{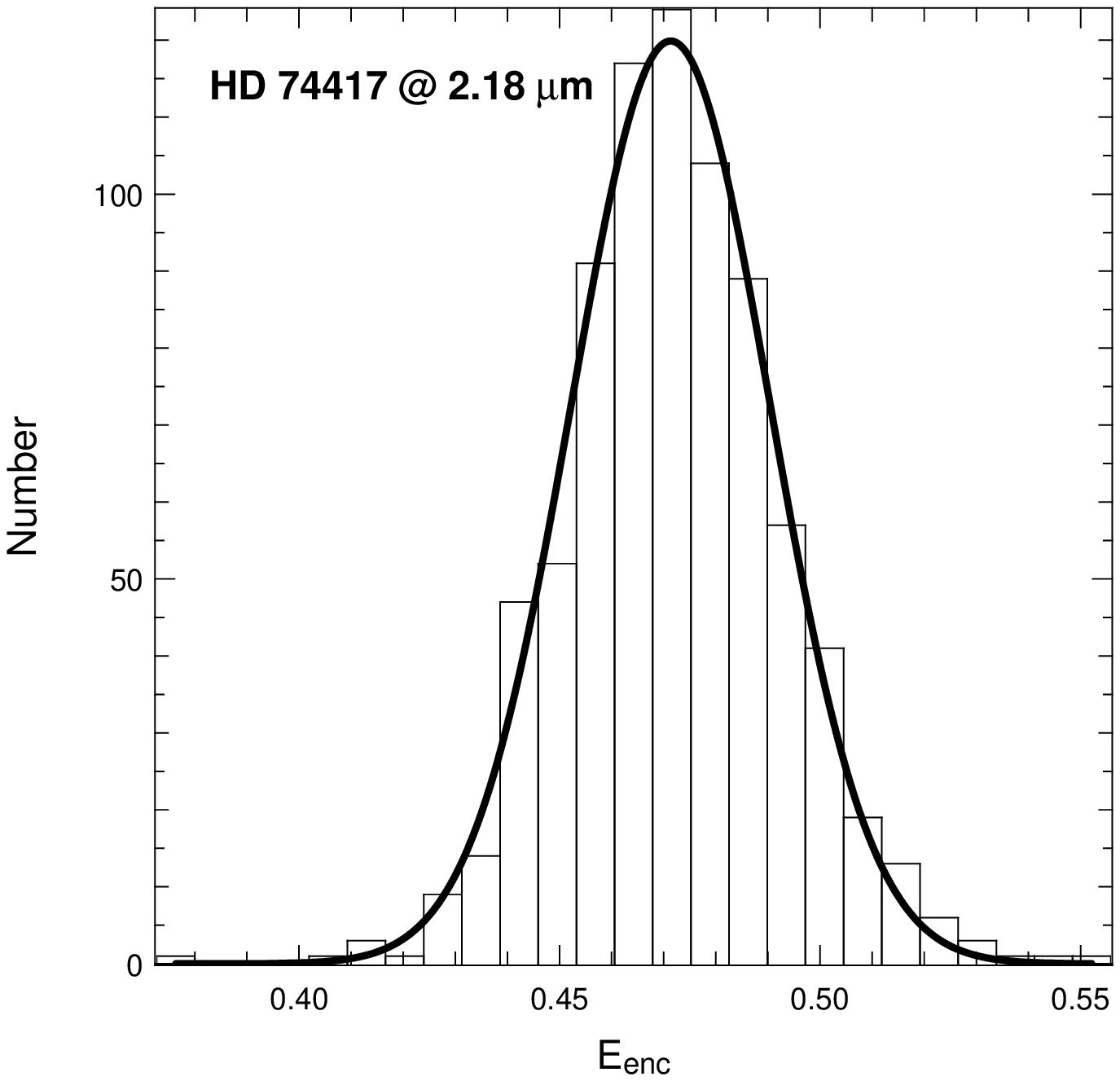}\hspace{1cm}
\includegraphics[width=4.45cm]{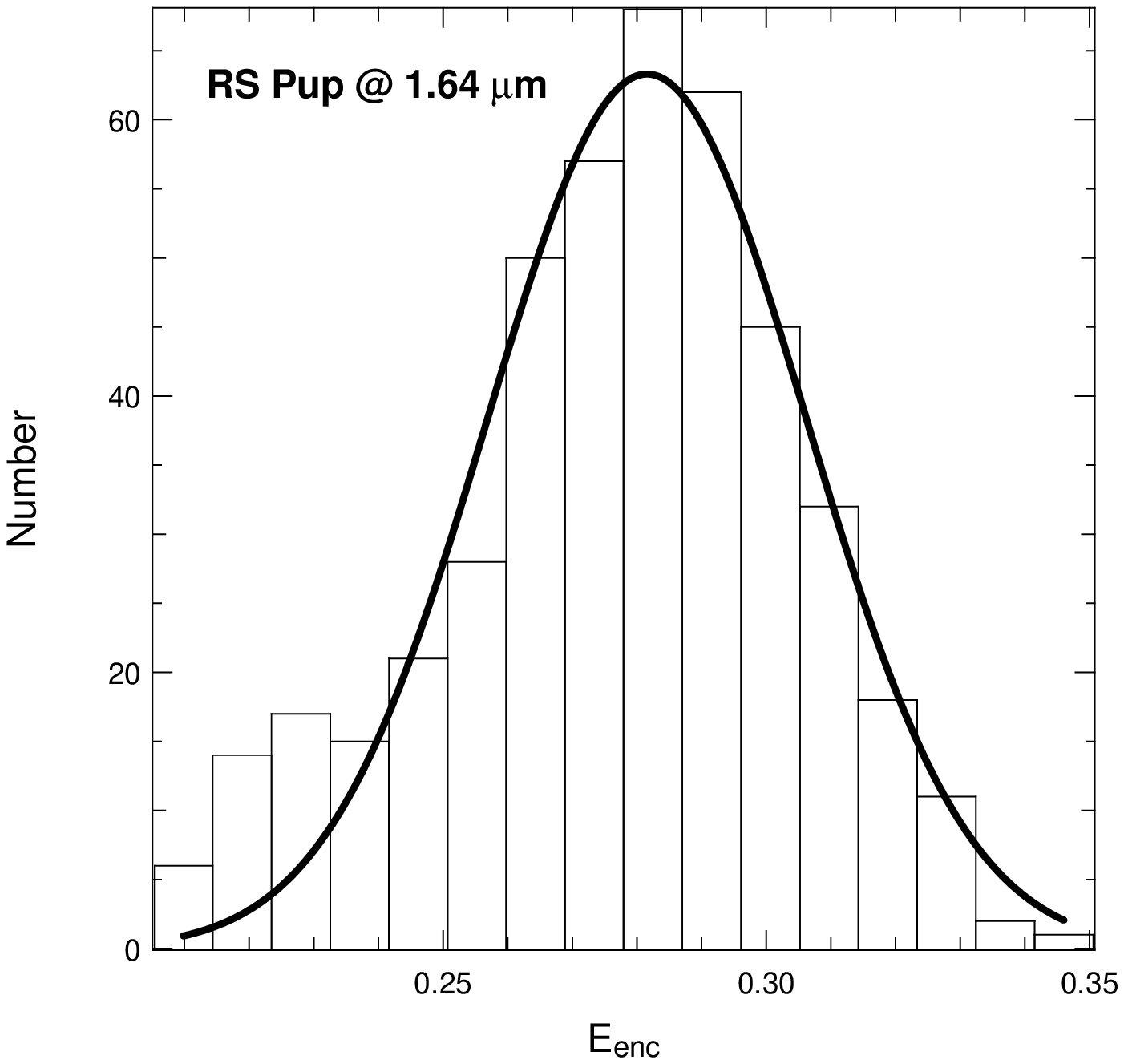}
\includegraphics[width=4.45cm]{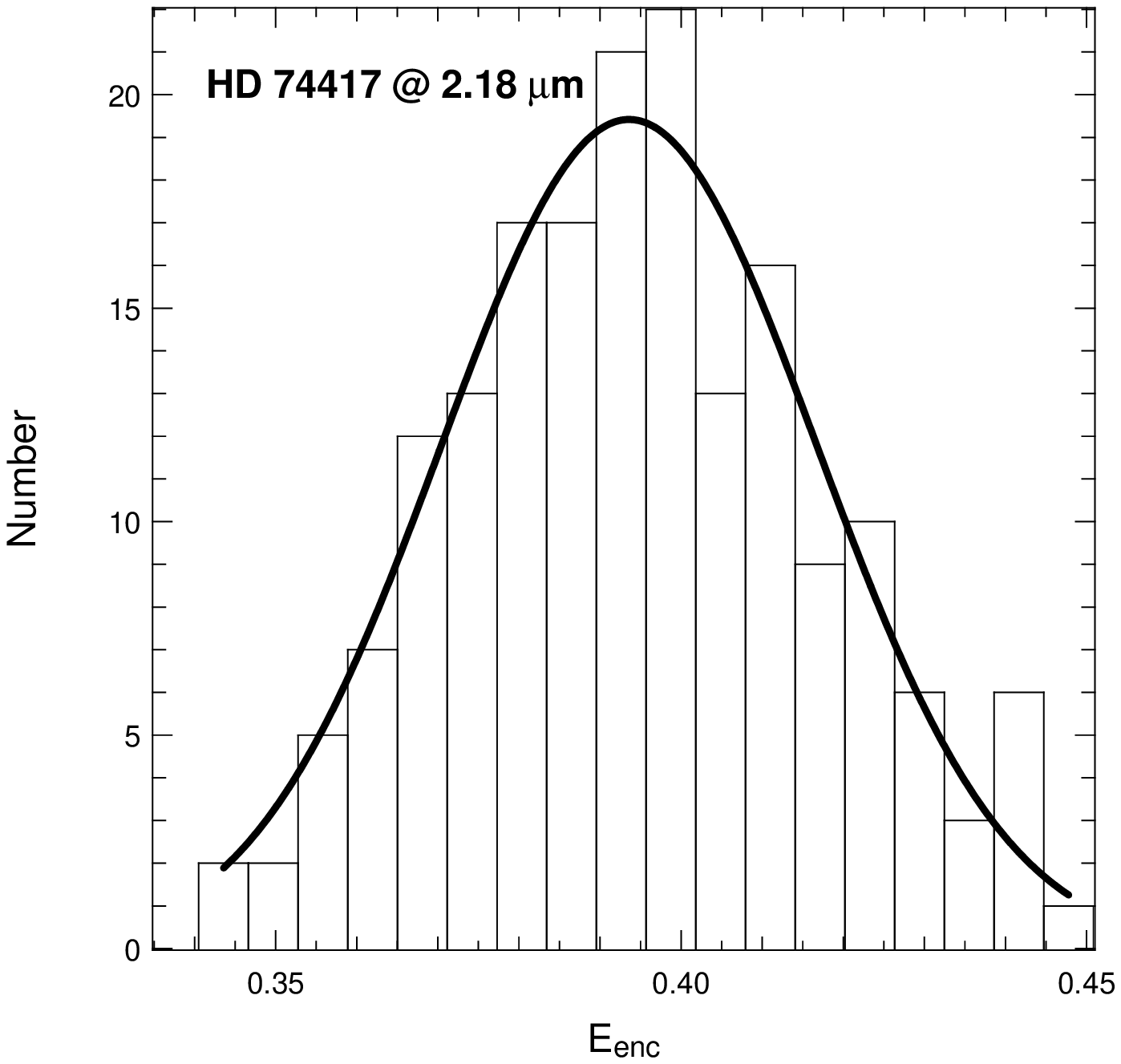}
\caption{Histogram of the encircled energy computed for 10\,\% best frames. The solid lines represent a Gaussian fit.}
\label{histogramme}
\end{figure}

\begin{figure*}[]
\centering
\includegraphics[width=6.6cm]{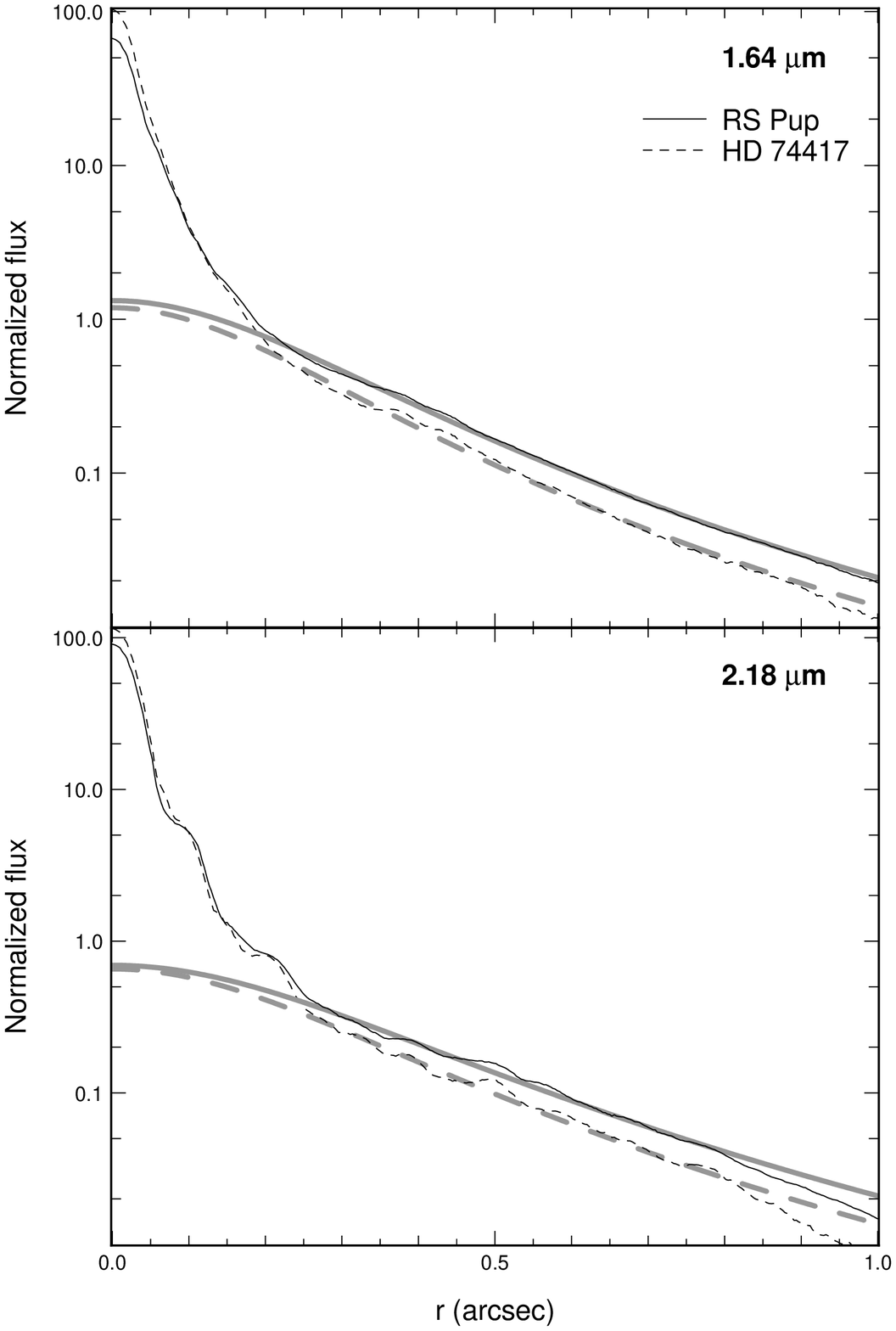}\hspace{.5cm}
\includegraphics[width=6.6cm]{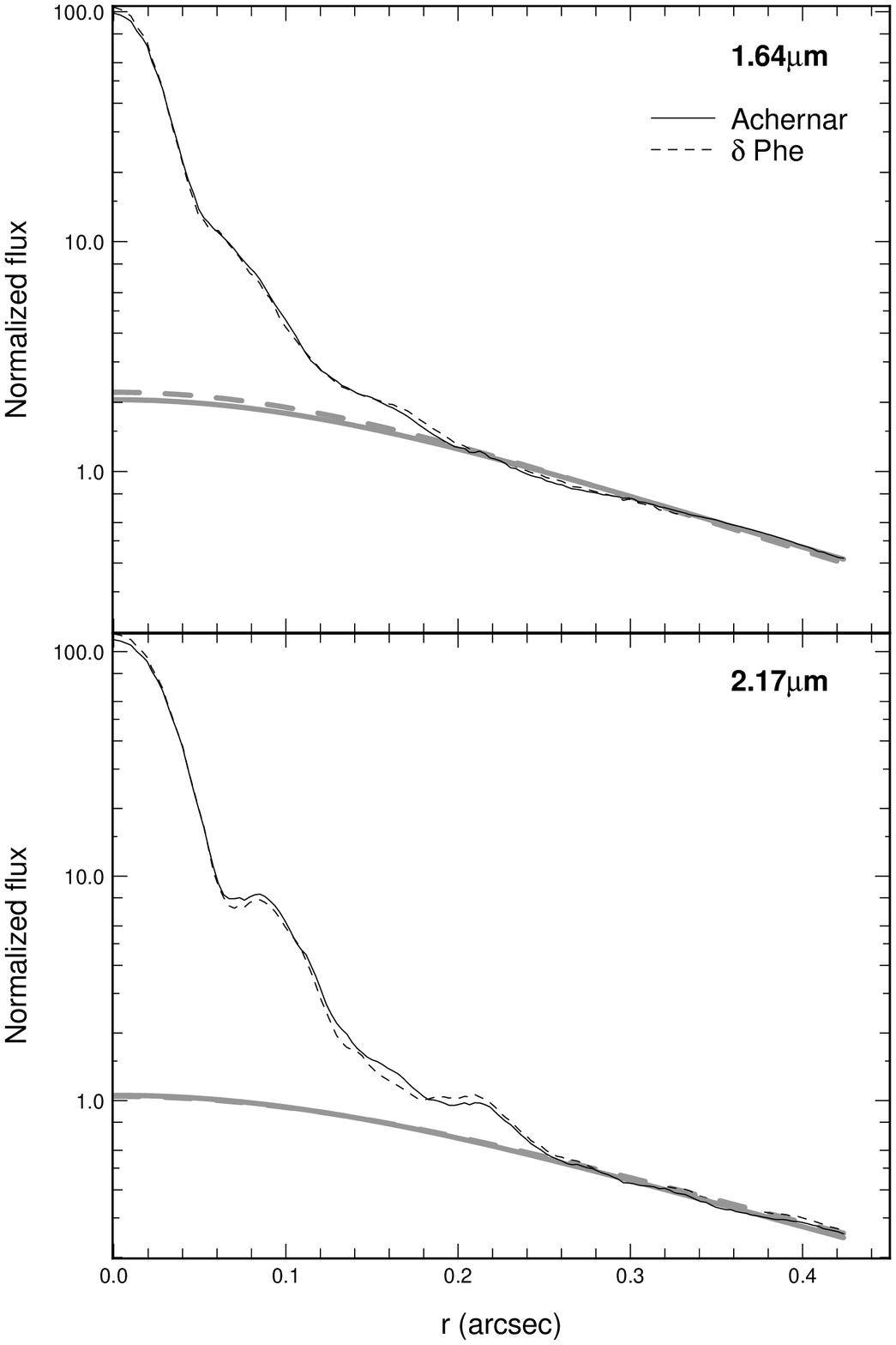}
\caption{Halo fit for all filters with logarithmic ordinates. The fluxes are normalized to the total flux. \textit{Left panel}: halo fit from the radial profile of images in the two filters for RS~Pup. \textit{Right panel}: halo fit from the radial profile of images in the two filters for Achernar. The fit in IB\_2.18 was for $r > 0.26\arcsec$ and in NB\_1.64 for $r > 0.2\arcsec$ for RS~Pup and for $r > 0.26\arcsec$ for Achernar. The dashed lines denote the reference stars, while the thick lines denote the Cepheid and Achernar. The fitted curves are plotted in gray.}
\label{halo_fit}
\end{figure*}

We show the resulting fits in Fig.~\ref{halo_fit}. They were applied for $r > 0.26$\arcsec at 2.18\,$\mu$m and 2.17\,$\mu$m while the range was for $r > 0.2$\arcsec at 1.64\,$\mu$m for RS~Pup (and its calibrator) and $r > 0.26$\arcsec for Achernar (and its calibrator). The black lines are our data (ring-median), while the gray lines represent the adjusted curves. The vertical scale is logarithmic and normalized to the total flux. We also evaluated the normalized encircled energy ($E\mathrm{_{enc}}$), i.e. the ratio of the flux in the core (i.e within 1.22$\mathrm{\lambda/D}$) to the total flux. We can already see from these curves that the flux in the halo is different in the images of RS~Pup and Achernar. These differences can be linked to the emission of the CSE around RS~Pup or a modification of the seeing between the two observations. This point will be discussed bellow with a statistical study of the encircled energy.

For RS~Pup in the NB\_1.64 filter we found from our fit comparing the two FWHMs, i.e ($\mathrm{\rho_{RSPup}-\rho_{HD74417}}$)/$\mathrm{\rho_{HD74417}}$, a change of halo size of 11\% and encircled energies $E\mathrm{_{enc}(RSPup)}$ = 0.327 and $E\mathrm{_{enc}(HD74417)}$ = 0.452 (see Table~\ref{fit_results_seeing}). This yields an encircled energy ratio $\varepsilon \sim$ 72\,\%.
The local DIMM station registered a seeing variation of around 26\,\%. Different values can be explained by the station location, which is not inside the telescope dome, but a few tens of meters below, rendering it more susceptible to ground layer effects that are known to affect the seeing \citep{Sarazin-2008-06}. As a result the UT seeing is consistently better than the one measured by the DIMM. For IB\_2.18 we show the fitting curves in Fig.~\ref{halo_fit} and the resulting fit parameters in Table~\ref{fit_results_seeing}. Our fit gives a seeing variation of around 9\,\%, while the DIMM measurements give a 29\,\% variation, leading to the same conclusion as before about the accuracy of the DIMM station seeing estimates. The encircled energies measured in IB\_2.18 are $E\mathrm{_{enc}(RSPup)}$ = 0.417 and $E\mathrm{_{enc}(HD74417)}$ = 0.517, giving an encircled energy ratio $\varepsilon\sim80\,\%$.

We evaluated the uncertainties according to the variations of the encircled energy in the cubes. Because we already selected the 10\,\% best frames, we kept these samples and plotted the histogram of the encircled energy to check the distribution function (Fig.~\ref{histogramme}). At the first order, the $E_{enc}$ distribution can be approximated by a Gaussian distribution. We then computed the relative standard deviation $\sigma(E\mathrm{_{enc}})/\overline{E_{enc}}$ for both stars. We assessed the uncertainty of the encircled energy ratio $\varepsilon$ with

\begin{equation}
\frac{\sigma(\varepsilon)}{\varepsilon} = \sqrt{ \left[ \frac{\sigma(E\mathrm{_{enc}})}{\overline{E\mathrm{_{enc}}}} \right]_{\mathrm{RSPup}}^2 + \left[ \frac{\sigma(E\mathrm{_{enc}})}{\overline{E_{\mathrm{enc}}}} \right]_{\mathrm{HD74417}}^2},
\label{uncertainties}
\end{equation}

We assume that the relative standard deviation of the encircled energy is a good approximation of the seeing variations.
We measured a relative standard deviation $[\sigma(E\mathrm{_{enc}})/\overline{E\mathrm{_{enc}}}]_{\mathrm{RSPup}}\sim8\,\%$ at 2.18\,$\mu$m and $\sim$\,10\,\% at 1.64\,$\mu$m while $[\sigma(E\mathrm{_{enc}})/\overline{E\mathrm{_{enc}}}]_{\mathrm{HD74417}}\sim$\,4\,\% and $\sim$\,6\,\% respectively. We also evaluated the statistical uncertainties with a bootstrap method that gave $\sigma\sim0.5\,\%$ for both stars. These statistical uncertainties do not take into account the seeing variations. We therefore chose Eq.~\ref{uncertainties} as a reliable estimate of the final uncertainty. This gives $\sigma(\varepsilon)/\varepsilon\sim9\,\%$ and $\sim12\,\%$ respectively in IB\_2.18 and NB\_1.64.

For Achernar, the fit gives a seeing variation $\sim3\%$ both in NB\_1.64 and IB\_2.17. The encircled energies measured in IB\_2.18 are $E\mathrm{_{enc}(Achernar)}$ = 0.495 and $E\mathrm{_{enc}(\delta\,Phe)}$ = 0.505, giving an encircled energy ratio $\varepsilon\sim98\,\%$ (see Table~\ref{fit_results_seeing}). In NB\_1.64 we have 0.374 and 0.376 respectively, which gives $\varepsilon\sim99\,\%$. The encircled energies are nearly identical, giving ratios close to 100\,\%. Using Eq.~\ref{uncertainties} as previously, we found a relative standard deviation $\sim15\,\%$ in IB\_2.17 and $\sim11\,\%$ in NB\_1.64. Because Achernar does not have an extended circumstellar component in NB\_1.64 and IB\_2.17, we attribute these variations to the Strehl variations and the AO correction quality. 
The encircled energy ratio seems stable between the two stars at the 2\,\% level in IB\_2.18 (see Table~\ref{fit_results_seeing}).

We can notice that the normalized curves of Achernar and its calibrator almost overlap, and this is what is expected for unresolved stars without envelope taken in same atmospheric and instrumental conditions. It is not the case for RS~Pup (see Fig.~\ref{radial_profil} and \ref{halo_fit}). The additional emission coming from the envelope of the Cepheid adds a component that is clearly not negligible. In the same atmospheric conditions and AO correction between the star and its calibrator, the encircled energy ratio should be close to 100\,\%. That is what we obtain from Achernar's data. Scaling RS~Pup's images to the same window size (i.e. $r\lesssim0.42\arcsec$) slightly changes $E\mathrm{_{enc}}$ because we normalized to the total flux, but we still have a discrepancy of 100\,\% for the encircled energy ratio ($\varepsilon\sim79\,\%$ and $\varepsilon\sim87\,\%$ respectively in NB\_1.64 and IB\_2.18). A smaller window on RS~Pup makes the total flux lower because we do not integrate all the envelope flux, consequently the normalized encircled energy is larger and therefore the $E\mathrm{_{enc}}$ ratio as well.


The drop in $E\mathrm{_{enc}}$ of RS~Pup compared with the statistics over Achernar's data and the fact that this drop is larger than the one directly estimated from the seeing halo variations, motivate the following section where this drop is analyzed under the assumption that it is caused by CSE emission.

\subsection{Encircled energy variation}
\label{encircled_energy_variation}

In this part we evaluate the flux of the envelope using the fact that the theoretical encircled energy ratio should be 100\% and that an additional component to the flux should decrease this ratio.

\begin{table}[]
\centering
\begin{tabular}{ccccccc} 
\hline
\hline
filter 		& name 							& $\rho$ (\arcsec)		& $f$							& E$\mathrm{_{enc}}$	\\
\hline
NB\_1.64 	& RS~Pup 							& 0.46 $\pm$ 0.01 		& 0.58 $\pm$ 0.01 		& 0.327						\\
                	& HD~74417						& 0.43 $\pm$ 0.01	 	& 0.43 $\pm$ 0.01 		& 0.452						\\
                	& Achernar 						& 0.49 $\pm$ 0.01 		& 1.00 $\pm$ 0.01 		& 0.374 						\\
                	& $\delta$~Phe 				& 0.46 $\pm$ 0.01		& 0.97 $\pm$ 0.01 		& 0.376						\\ \hline
IB\_2.18 	& RS~Pup 							& 0.56 $\pm$ 0.01 		& 0.45 $\pm$ 0.01 		& 0.417						\\
                	& HD~74417 					& 0.50 $\pm$ 0.01 		& 0.34 $\pm$ 0.01	 	& 0.517						\\ \hline
NB\_2.17 	& Achernar 						& 0.52 $\pm$ 0.01		& 0.59 $\pm$ 0.01		& 0.495 						\\
                	& $\delta$~Phe 				& 0.54 $\pm$ 0.01		& 0.61 $\pm$ 0.01 		& 0.505 						\\ \hline
\end{tabular}
\caption{Best-fit results for the residual halo where. $\rho$ is the FWHM and $f$ is a parameter proportional to the flux of the halo normalized to the total flux. $E\mathrm{_{enc}}$ represents the flux in the central core (i.e. within 1.22$\lambda/D$) to the total flux. Note: The fitted $I_{\mathrm{halo}}$ analytical function is not normalized to $f$, i.e $\int I_{\mathrm{halo}}rdr \neq f$.}
\label{fit_results_seeing}
\end{table}

The encircled energy of RS~Pup can be written as

\begin{displaymath}
E_{\mathrm{enc}}=\frac{F_{\mathrm{coh}} + \alpha F_{\mathrm{env}}}{F + F_{\mathrm{env}}},
\end{displaymath}
where $F$ is the total flux of the star, without the extended flux, $F\mathrm{_{coh}}$ is the flux in the coherent core, $F\mathrm{_{env}}$ is the emission of the environment and $\alpha$ is the fraction of the envelope flux that lays in the core. For simplicity, we will assume $\alpha << 1$, which means that we assume that the CSE is significantly larger than the core. Accordingly we can write

\begin{displaymath}
E_{\mathrm{enc}}\approx\frac{F_{\mathrm{coh}}/F }{1 + F_{\mathrm{env}}/F},
\end{displaymath}

Assuming that the encircled energy is stable in NACO for our observing conditions, we can calibrate the previous formula to obtain the extended emission flux:

\begin{displaymath}
\varepsilon = \frac{E\mathrm{_{enc}(Cepheid)}}{E\mathrm{_{enc}(reference)}}\approx\frac{1}{1 + F(\mathrm{env})/F(\mathrm{Cepheid})}
\end{displaymath}

The relative emission $F(\mathrm{env})/F(\mathrm{Cepheid})$ can be estimated with the encircled energy ratio from our measured values (Table~\ref{fit_results_seeing}). We can also take the uncertainty values assessed in the previous section, defined as being the relative standard deviation on 10\% of best frames in the cubes. We found a relative extended emission around RS~Pup contributing to 38\,$\pm$\,17\,\% of the star's flux in the NB\_1.64 band and 24\,$\pm$\,11\,\% in IB\_2.18.


\section{Morphological analysis}
\label{morphological-analysis}

We now present a qualitative analysis of the envelope morphology. In this analysis, based on the statistical properties of the noise owing to speckle boiling, we presented the ratio $\sigma(F)/\overline{F}$ as a function of the factor ($g$). This parameter is intrinsic to speckle temporal and spatial structures, is not expected to depend on azimuth, and is proportional to the envelope flux. We mapped this factor, and any azimuthal asymmetry should be attributed to the morphology of the envelope.

\subsection{Analysis method}

The principle of the method is to recover the flux of the envelope in the short exposures.
The difference between a simple unresolved star and a star surrounded by an envelope is that the envelope will prevent the flux between the speckles from reaching zero.

Statistical descriptions of stellar speckles have been proposed by many authors \citep{Racine-1999-05,Fitzgerald-2006-01}, who showed that speckle noise is not negligible. The distribution of speckle statistics in AO were also studied by \citet{Canales-1999-}. For example, \citet{Racine-1999-05} (hereafter Ra99) showed that the speckle noise dominates other noise sources within the halo of bright stars corrected with AO systems. We know that the total variance in the cube short exposures after the basic corrections (i.e corrected for bias, bad pixels, flat-field and recentered) is

\begin{equation}
\sigma^2(r) = \sigma_\mathrm{s}^2(r) + \sigma_{\mathrm{ph}}^2(r) + \sigma_{\mathrm{RON}}^2 + \sigma^2_{\mathrm{Strehl}}(r),
\label{variance-noise}
\end{equation}
where $\sigma_\mathrm{s}^2$ is the variance of the speckle noise, $\sigma_\mathrm{{ph}}^2$ the variance of the photon noise, $\sigma_\mathrm{{RON}}^2$ the variance of the readout noise and $\sigma_{\mathrm{Strehl}}^2$ the noise owing to Strehl ratio variations. We neglected the variance of the sky photon noise, which is negligible in our wavelength for bright targets. 
Because the Strehl ratio variations occur in the coherent core and because we are particularly interested in the halo part, we omit these variations in the rest of this section.

\begin{figure*}[ht]
\centering
\includegraphics[width=6.6cm]{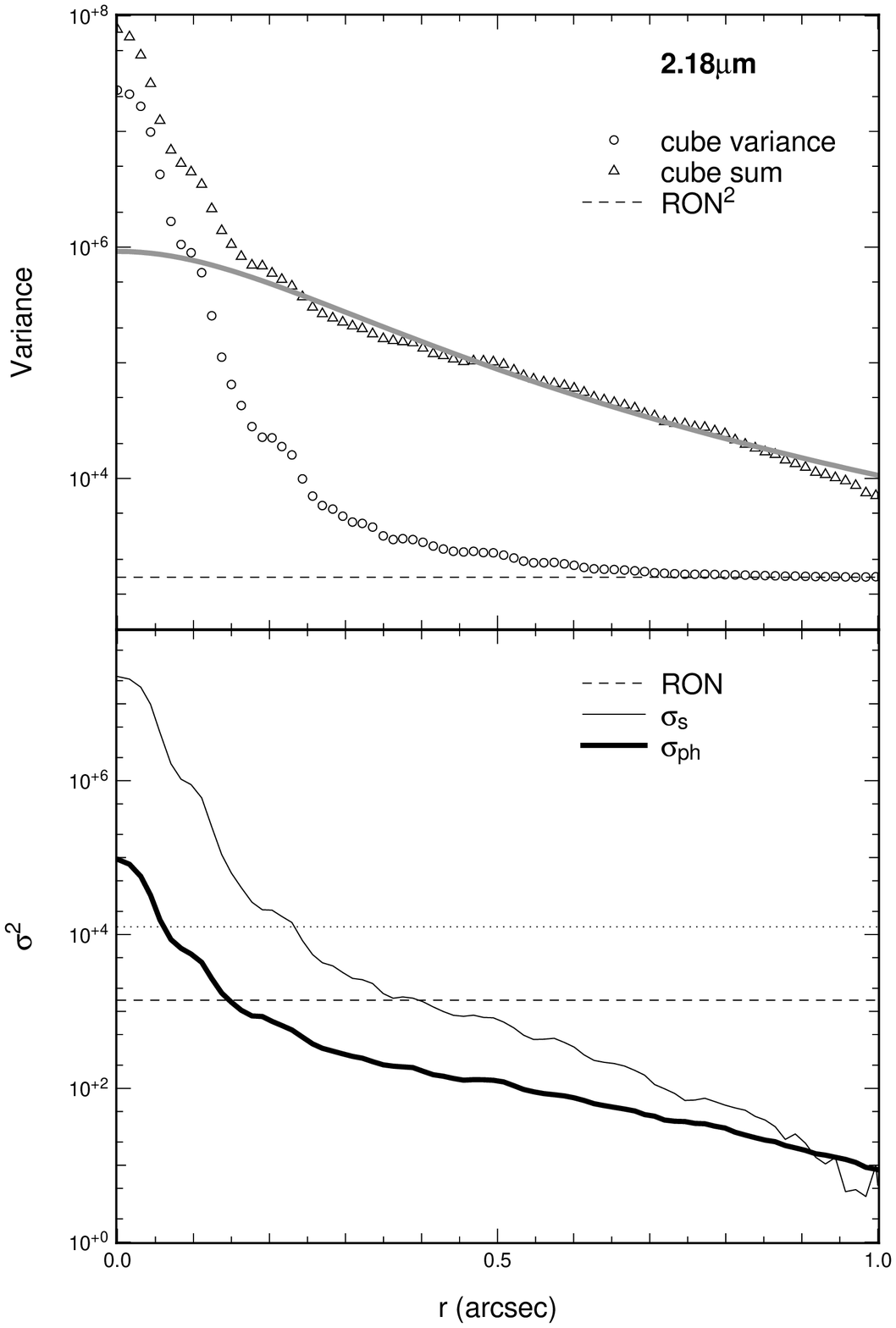}\hspace{.5cm}
\includegraphics[width=6.6cm]{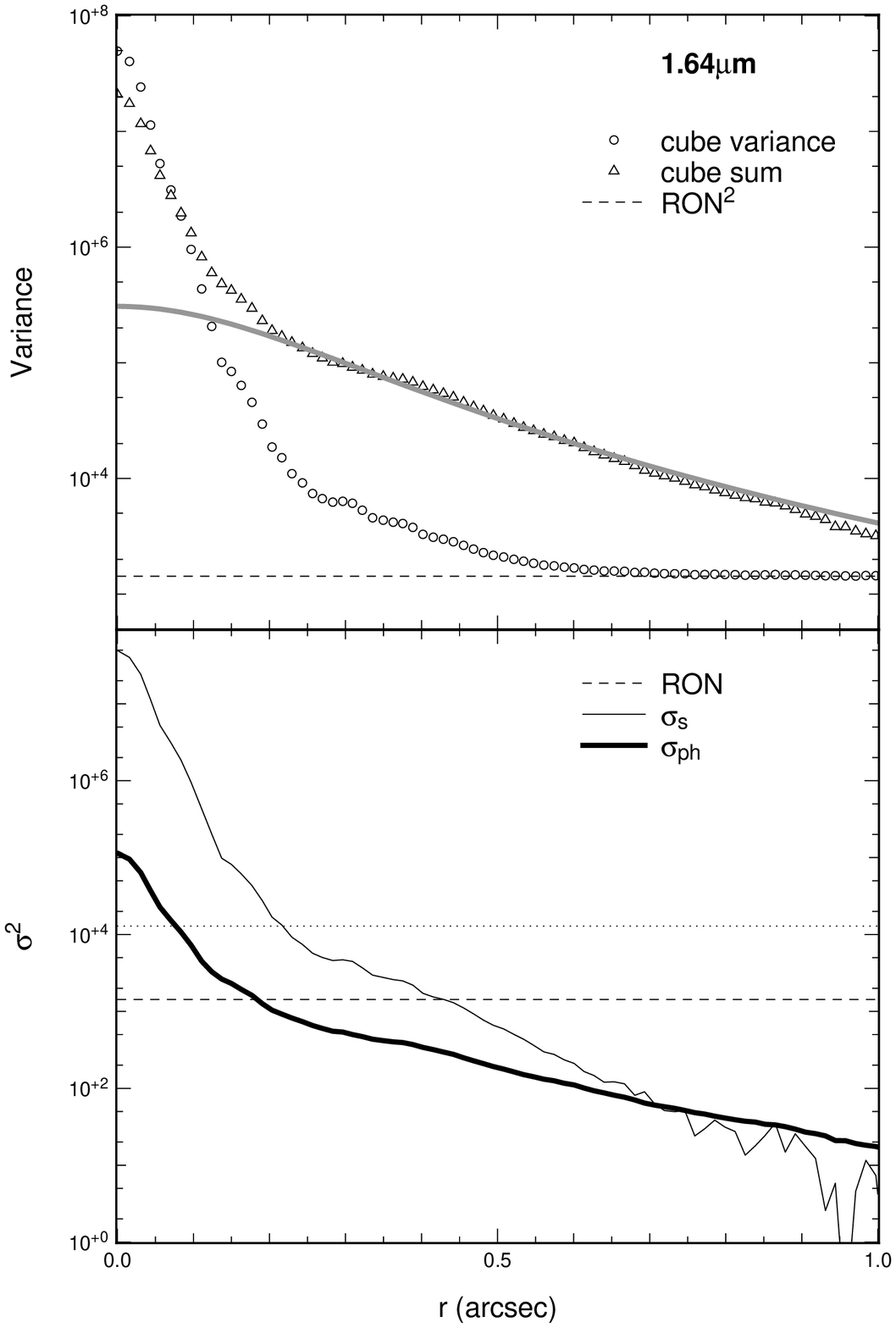}
\caption{\textit{Left panel}: at the top we plot the variance (circles) and the average value (triangle) of one data cube of HD~74417 at 2.18\,$\mu$m. \textit{Bottom}: contribution of the different noise sources. \textit{Right panel}: the same for the 1.64\,$\mu$m filter. The scale for all plots is logarithmic in ordinates. The dotted curves show the 3$\mathrm{\sigma}$ limits of the readout noise, while the gray curves denote the halo fit.}
\label{variance_1}
\end{figure*}

We present for one calibrating cube the variance observed in the top panels of Fig.~\ref{variance_1} for both filters. We plotted the radial profile computed using median rings on all azimuths. The dashed curve denotes the variance of the readout noise. We can estimate some of the components in Eq.~\ref{variance-noise}  from our data. For example, from the stacked image that we obtained by summing the cube, we obtain the core and halo profile, as we did in Sect.~\ref{shiftandadd}. We then obtain the photon noise variance using its Poissonian property, knowing that the variance is equal to the mean. We define for the stacked image the radial intensity profile $I(r)$ and we estimate the term $\sigma_\mathrm{{ph}}^2$ of Eq.~\ref{variance-noise} with

\begin{displaymath}
\sigma_{\mathrm{ph}}^2(r) = \gamma\ I(r)  = \gamma\ \frac{I(r)}{N},
\end{displaymath}

\noindent where $N$ is the number of frames in the cube (1999 frames in IB\_2.18 and 457 in NB\_1.64) and $\gamma$ the gain of the detector in order to convert noises in electron unit. The readout noise is estimated from the outer part of the images (see in Fig.~\ref{variance_1} the top panels). We estimate the speckle noise from the total variance of the cubes as

\begin{displaymath}
\sigma_\mathrm{s}^2(r) = \sigma^2(r) - \sigma_{\mathrm{ph}}^2(r) - \sigma_{\mathrm{RON}}^2,
\end{displaymath}



\begin{figure*}[!ht]
\centering
\includegraphics[width=6.6cm]{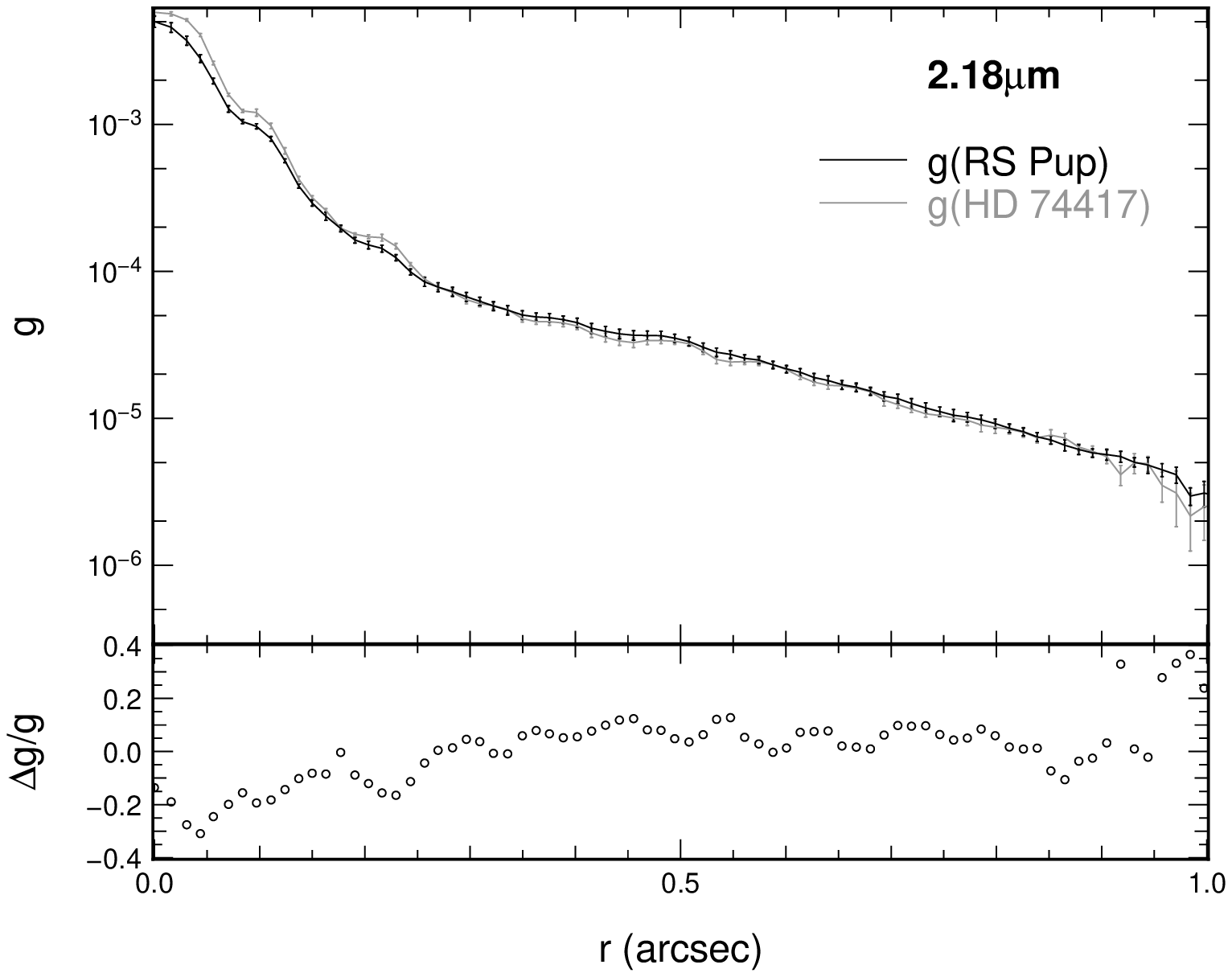}\hspace{.5cm}
\includegraphics[width=6.6cm]{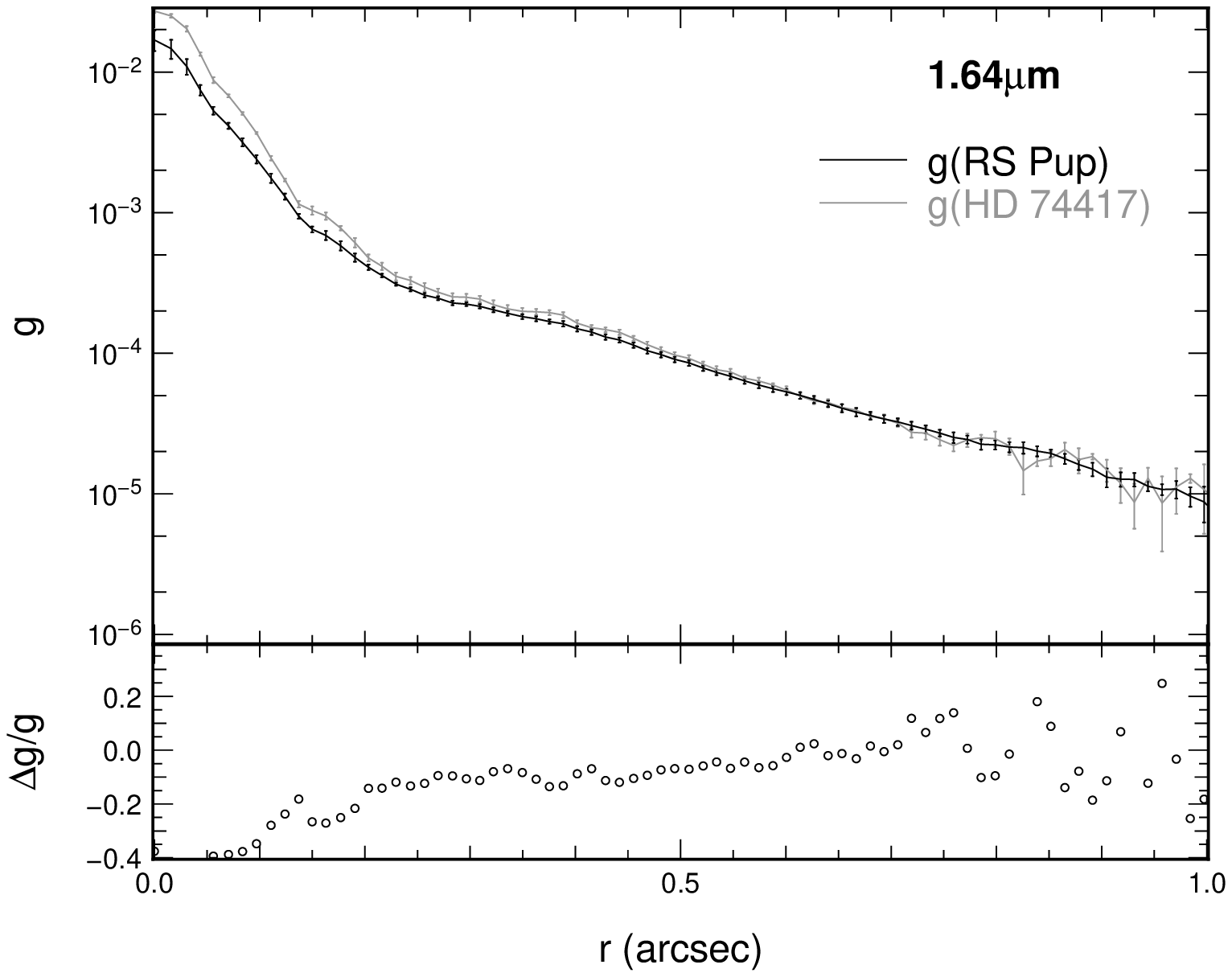}
\caption{Mean $g$ function for each star. The error bars correspond to the RMS. The bottom plot represents the absolute relative difference. We can see a larger difference in the core owing to the Strehl variations. From $r > r_1$ the functions are similar at a 9\,\% level in NB\_1.64 and at 6\,\% in IB\_2.18.}
\label{invariant}
\end{figure*}

In the bottom panels of Fig.~\ref{variance_1} we present the relative importance of the different noise sources. We see that the different variances show a different radial behavior, and we see in particular that the photon noise of the core never dominates. More precisely, we can define two regimes:

\begin{itemize} 
\item $\sigma(r) \simeq \sigma_\mathrm{s}(r)$ for $ r \lesssim r_\mathrm{1}$, in this regime, the speckle noise dominates the other variances ($\sigma_\mathrm{s} > 3\,\sigma_{\mathrm{RON}}$)
\item $\sigma(r) \simeq \sqrt{ \sigma_\mathrm{s}^2(r) + \sigma_{\mathrm{RON}}^2} $ for $r \gtrsim r_\mathrm{1} $
\end{itemize}

\noindent with $r_\mathrm{1} \simeq 0.22\arcsec$ in NB\_1.64 band and $r_\mathrm{1} \simeq 0.24\arcsec$ in IB\_2.18. The fairly weak ($\propto \lambda^{-12/5}$) wavelength dependency of the speckle noise may explain the same order of magnitude between the limit radius in IB\_2.18 and NB\_1.64.

Indeed there is another regime in the range $0 \leqslant r < r_\mathrm{1}$ where the Strehl variations occur. The PSF core is mainly affected by these variations. As said previously, we are interested in the PSF halo to detect the envelope between the speckles, which is why we neglect this term.

\begin{figure*}[ht]
\centering
\includegraphics[width=6.6cm]{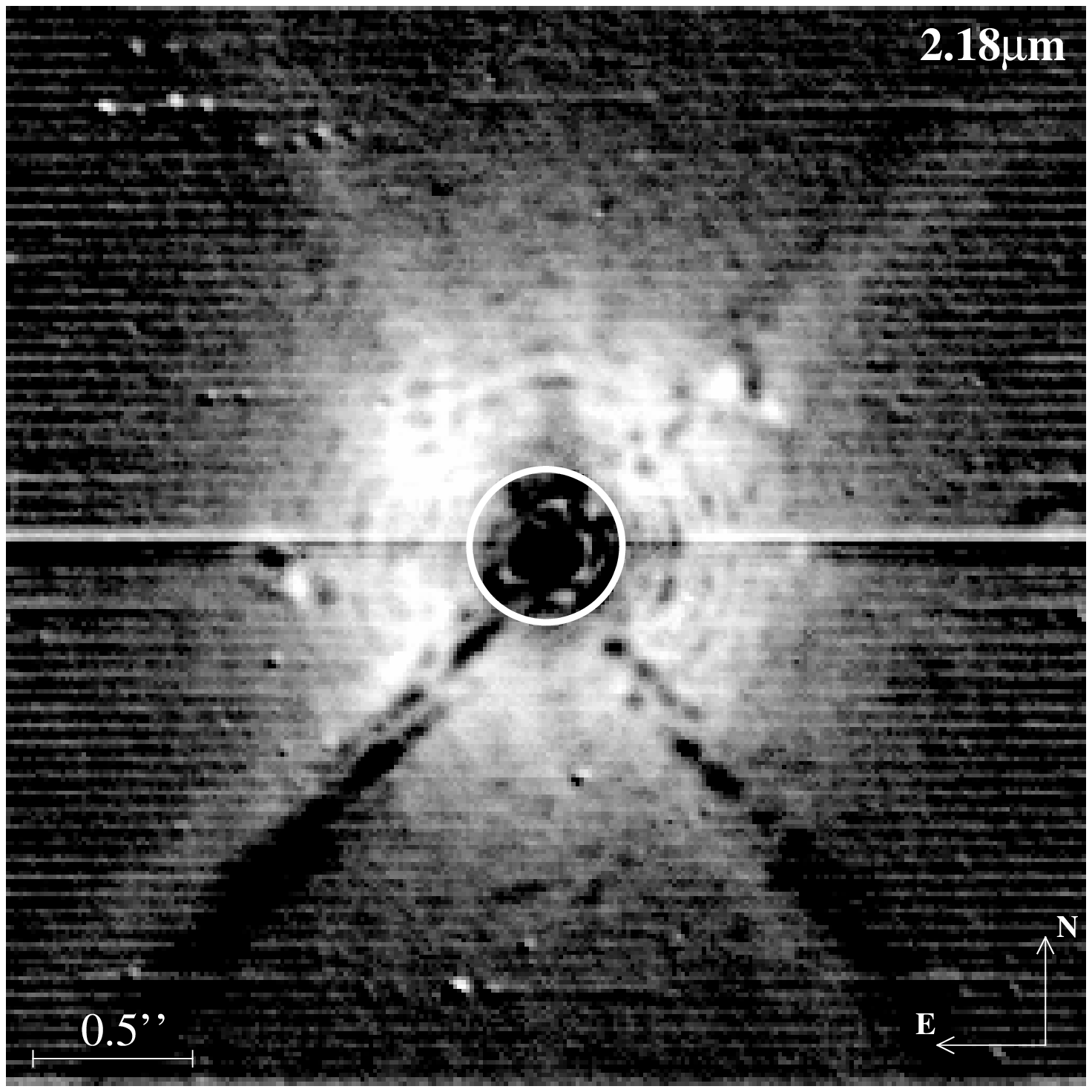}\hspace{.5cm}
\includegraphics[width=6.6cm]{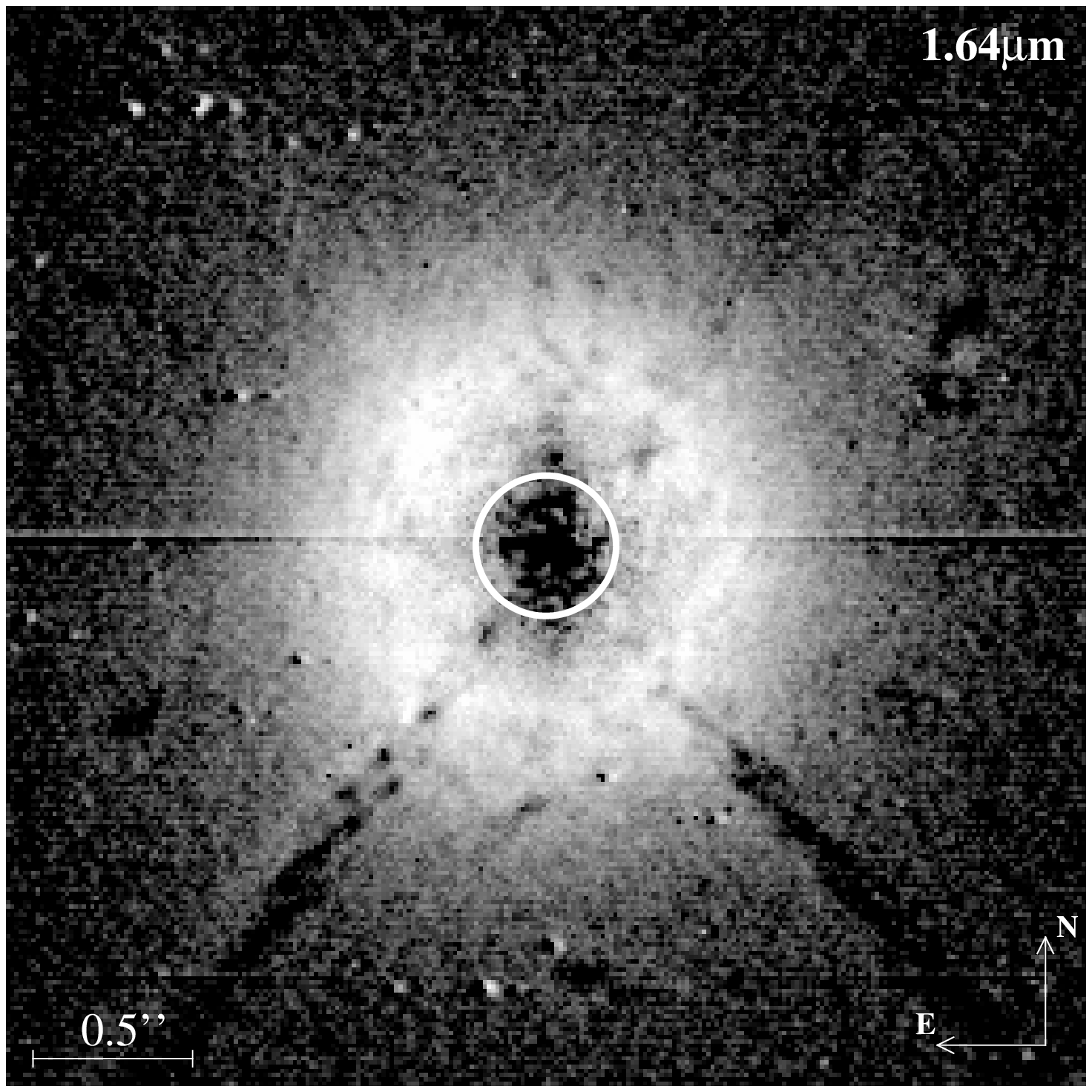}
\caption{$\xi$ parameter (Eq.~\ref{region_2}) proportional to the envelope flux at 2.18\,$\mu$m (\textit{left}) and at 1.64\,$\mu$m (\textit{right}). The white circle represents the $r_1$ limit of Eq.~\ref{env} separating the two regimes. The scale intensity is logarithmic.}
\label{env}
\end{figure*}

\subsection{Speckle noise invariant}

We use an interesting property of the speckle noise, namely that its variance is proportional to the squared flux in the cube (Ra99), $F/\sigma$ is therefore an invariant. More precisely:

\begin{displaymath}
\sigma_\mathrm{s} = g(r,\lambda,\Delta t,S,...)\,F_\star (\lambda),
\end{displaymath}

\noindent where the function $g$ is a function depending on the wavelength $\lambda$, the integration time $\Delta t$, the Strehl ratio $S$ and atmospheric parameters. 

We checked this property by plotting in Fig.~\ref{invariant} the mean $g$ function for each star, i.e we evaluated a $g$ function for each cube that we then averaged (the error bars correspond to the standard deviation between the cubes). We took care to remove the flux contribution from the circumstellar envelope with the values found in Sect.~\ref{shiftandadd}. Obviously there is no clear difference between the two stars except in the core. This can be explained by the Strehl ratio variations we omitted before in the core.
We can assess these variations by evaluating the encircled energy for each frame of each calibrating cube (here we take 100\,\% of frames, without selection) and by using it as an estimator of the Strehl ratio in function of time. By fitting a Gaussian distribution to $E_{\mathrm{enc}}$ for each star and both filters, we estimate the relative standard deviation for RS~Pup to be $\sigma(E_{\mathrm{enc}})/\overline{E_\mathrm{{enc}}} \sim 15\,\%$ in IB\_2.18 and $\sim 17\,\%$ in NB\_1.64. For HD~74417 we found $\sim\,8\,\%$ and $\sim\,9\,\%$ respectively at 2.18\,$\mu$m and 1.64\,$\mu$m.

The total invariance of the $g$ function, i.e for the complete radius, depends on the Strehl ratio variations. Nevertheless, because these variations occur in the coherent core, we can neglect them if we go far away from the central part. For $r > r_\mathrm{1}$ we found a mean absolute relative difference (see Fig.~\ref{invariant}) of  5\,\% in IB\_2.18 and 8\,\% in NB\_1.64. So for $r > r_1$ the invariance of the $g$ function is verified with a good accuracy.

We can measure $g$ using our reference star. Assuming it is stable in time, we can define the observable $\Gamma$ as the average of the flux of a given pixel in the cube, divided by the standard deviation. Defining $F_{\mathrm{cal}} = \alpha F_{\mathrm{RS~Pup}} = \alpha F_\star$ and making the difference between $\Gamma_{\mathrm{sci}}$ and $\Gamma_{\mathrm{cal}}$ (where the index sci and cal refer to RS~Pup and HD~74417 respectively) we have

\begin{equation}
\xi = \Gamma_{\mathrm{sci}} -\Gamma_{\mathrm{cal}}= \frac{F_\star + F_{\mathrm{env}}}{ \sqrt{ g_\star^2  F_\star^2 + \sigma_{\mathrm{RON}}^2 } } - \frac{\alpha F_\star}{ \sqrt{ g_\star^2 \alpha^2 F_\star^2 + \sigma_{\mathrm{RON}}^2}},
\label{gamma-diff}
\end{equation}

All variables in Eq.~\ref{gamma-diff} are functions of the pixel coordinates $(x,y)$, but we omit this notation for clarity. 

The parameter $\alpha$ can be estimated with the results from Sect.~\ref{shiftandadd}. The flux ratio between the two final images of RS~Pup and HD~74417 is a function of $\alpha$ and $F(\mathrm{env})/F_\star$:

\begin{equation}
\frac{F(\mathrm{RS~Pup})}{F(\mathrm{HD~74417})} = \frac{F_\star + F_\mathrm{env}}{\alpha\ F_\star} = \frac{ 1 + F_\mathrm{env} / F_\star } {\alpha},
\label{region_2}
\end{equation}

By measuring the total flux in our final images (Sect.~\ref{shiftandadd}) and using $F_\mathrm{env} / F_\star$ estimated in Sect.~\ref{encircled_energy_variation} we get $\alpha = 1.06\,\pm\,0.13$ at $1.64\,\mu m$ and $\alpha = 0.96\,\pm\,0.08$ at $2.18\,\mu m$. We also used the total flux variations in our 10 (for RS~Pup) and 4 (for the reference) averaged images to assess the uncertainties, but they turn out to be small ($< 2\,\%$) compared to $F_\mathrm{env} / F_\star$ uncertainties. As a first approximation we can simplify Eq.~\ref{gamma-diff} with $\alpha \sim 1$ to obtain

\begin{equation}
\xi = \frac{F_\mathrm{{env}}}{ \sqrt{ g_\star^2  F_\star^2 + \sigma_{\mathrm{RON}}^2 }}.
\label{region_2}
\end{equation}

$\xi$ is thus directly proportional to the flux of the CSE. We expect this function to be null if there is no envelope. The denominator is a radial parameter, therefore any spatial asymmetry in $\xi$ will be linked to an asymmetry of the CSE.

From our different data cubes we computed a mean $\xi$ by subtracting to each $\Gamma_{\mathrm{sci}}$ a mean $\Gamma_{\mathrm{cal}}$, i.e:

\begin{displaymath}
\overline{\xi(x,y)} = \frac{1}{n} \sum_{j=1}^n ( \Gamma_{\mathrm{sci, j}}  - \overline{\Gamma_{\mathrm{cal}}} )\ \ \mathrm{and}\ \ \overline{\Gamma_{\mathrm{cal}}} = \frac{1}{m} \sum_{i=1}^m \Gamma_{\mathrm{cal, i}},
\end{displaymath}

\noindent where $n$ and $m$ are the number of cube for RS~Pup and HD~74417 respectively. This process provides us with an average image proportional to the flux of the CSE in each filter.


\subsection{Symmetry of the CSE}
\label{shape-of-the-cse}

In Fig.~\ref{env} we show the $\xi$ function for the NB\_1.64 and IB\_2.18 band. For radii larger than $r_1$, we expect the function to be null if there is no CSE (see Eq.~\ref{region_2}), whereas we clearly obtain a positive image, indicative of a likely detection of a CSE. The central part of the image with $r < r_1$ (white circle for $r = r_1$) denotes the part where the previously defined function $g$ is not invariant because of Strehl ratio fluctuations. It is difficult to estimate and therefore does not supply reliable informations close to the star.

\begin{figure}[h]
\centering
\includegraphics[width=4.45cm]{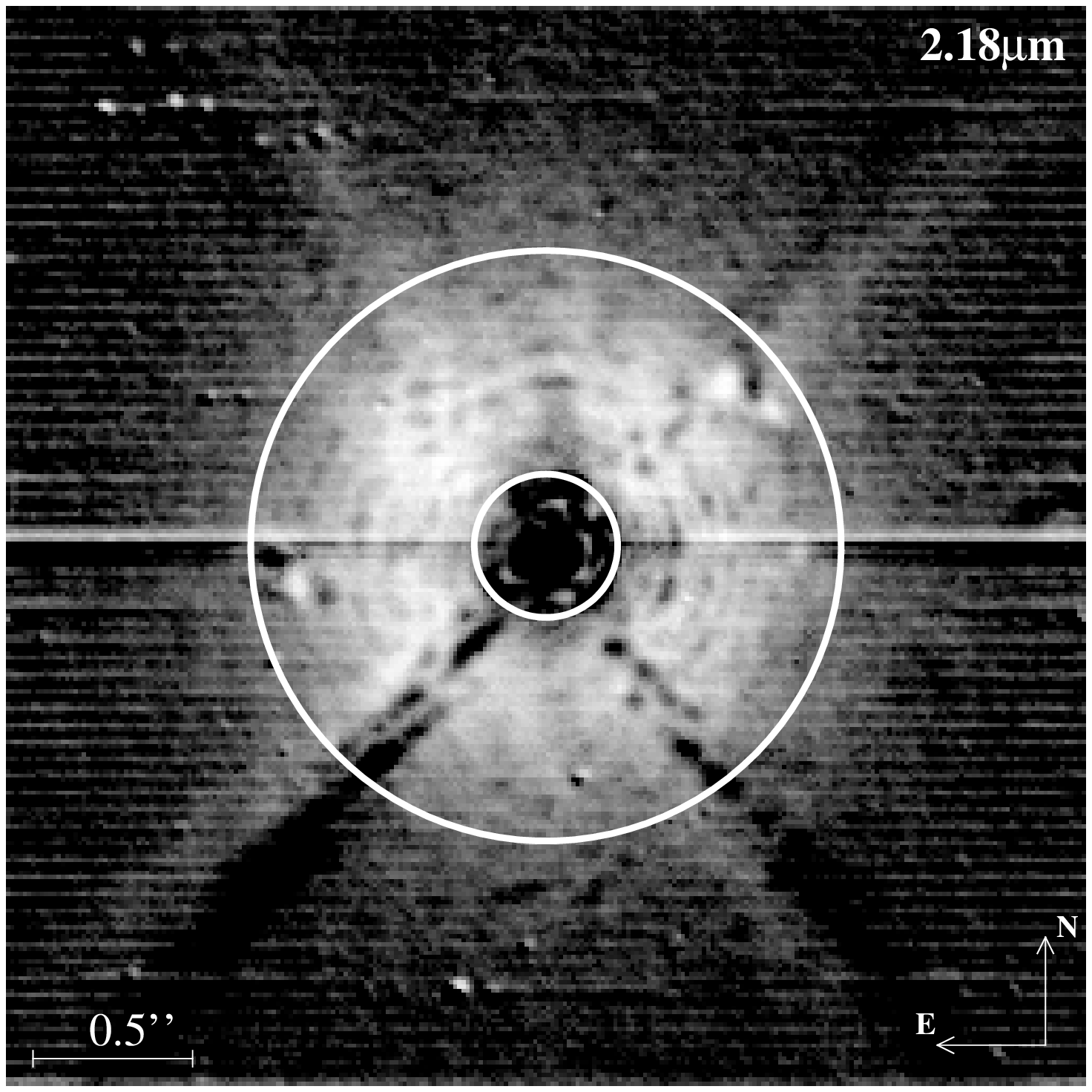}
\includegraphics[width=4.45cm]{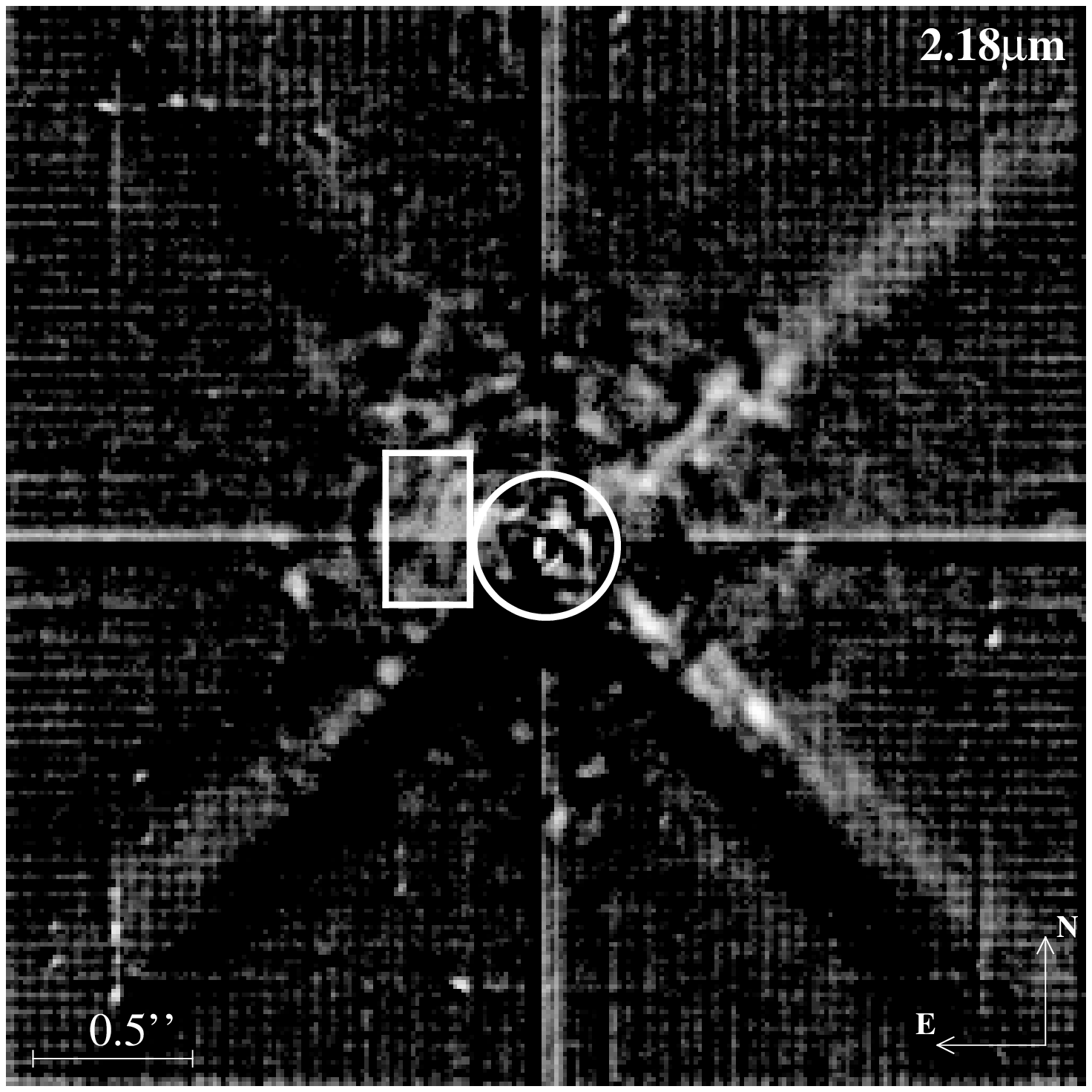}\hspace{1cm}
\includegraphics[width=4.45cm]{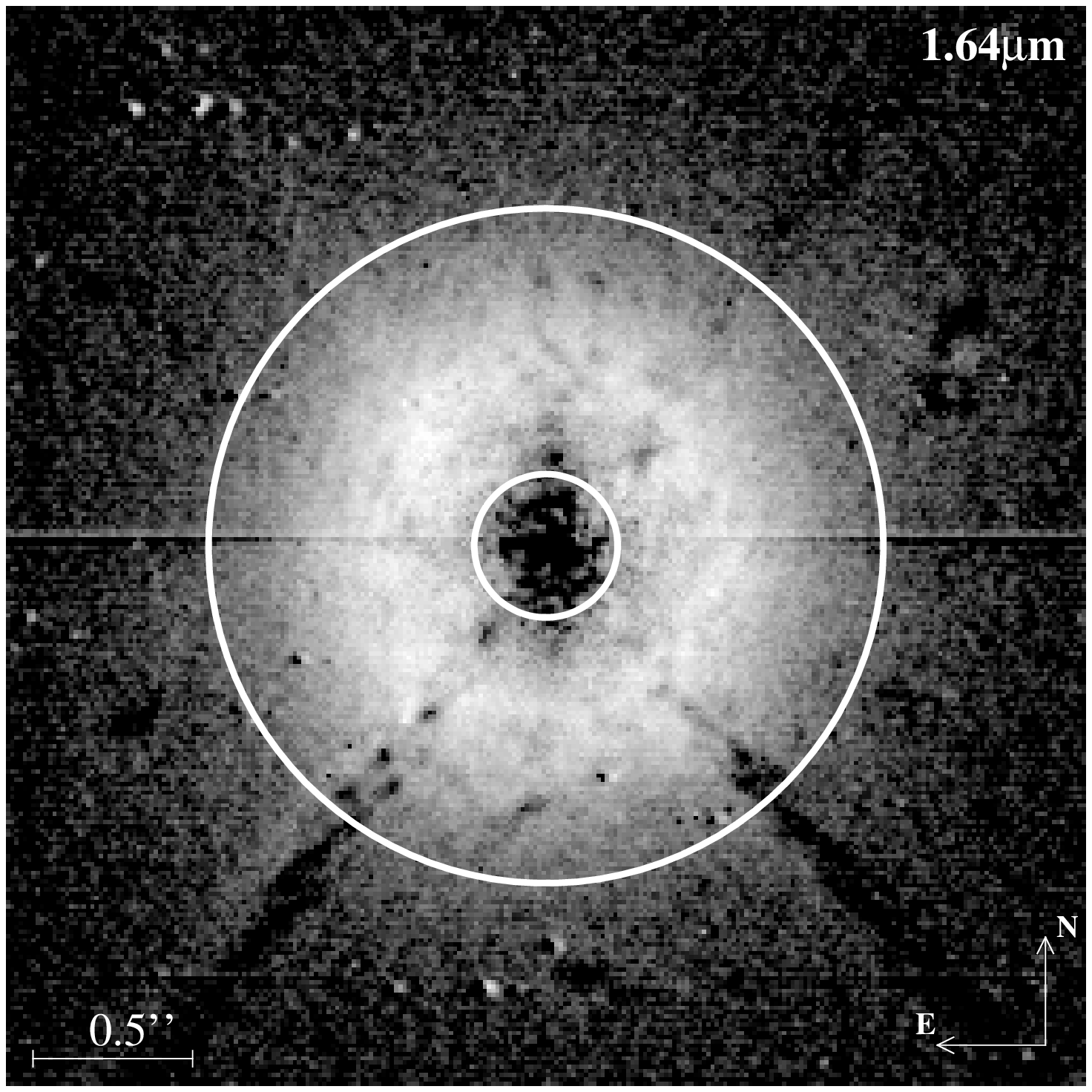}
\includegraphics[width=4.45cm]{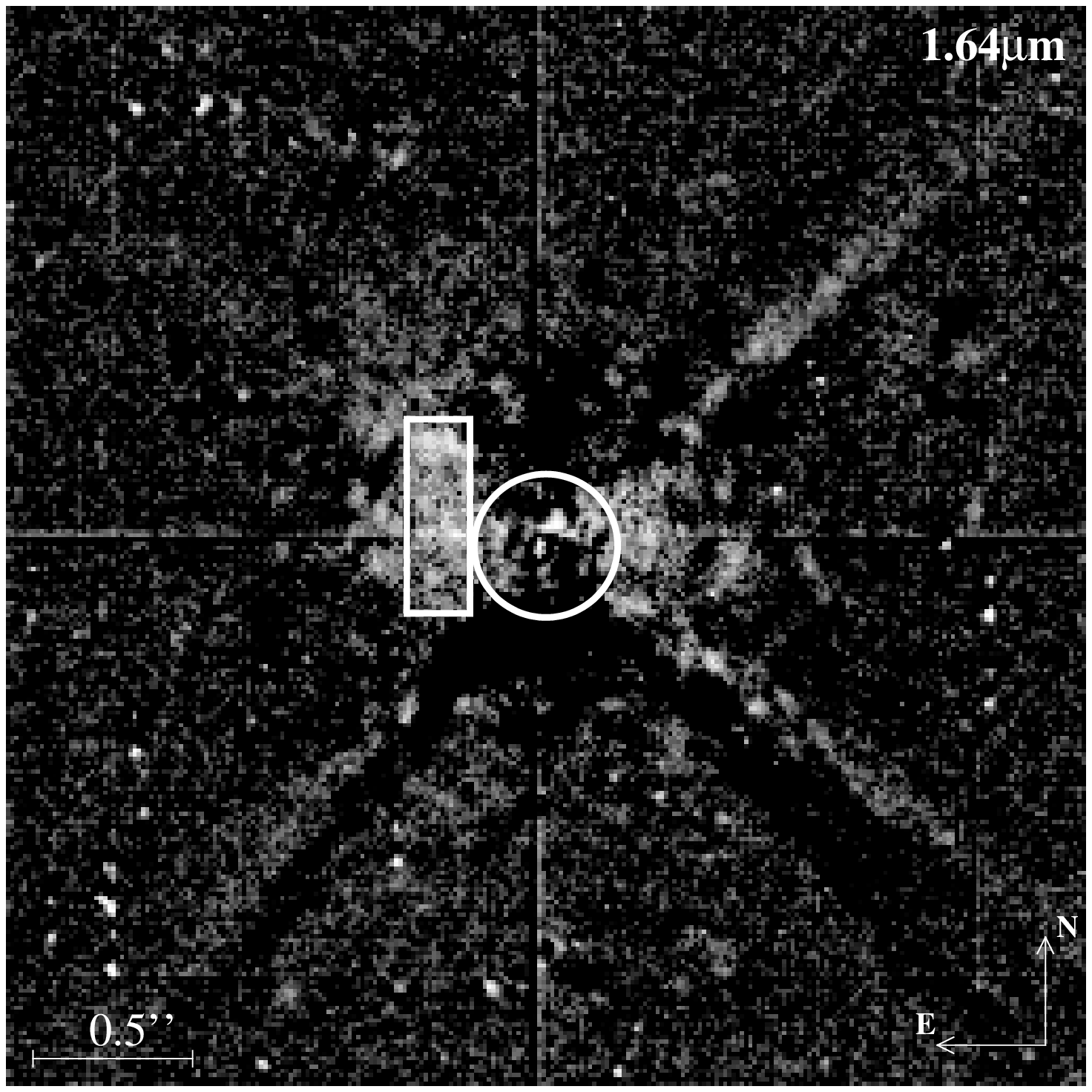}
\caption{Estimate of the symmetry level. \textit{Top panels:} on the left is plotted the $\xi$ parameter at $2.18\,\mu m$ with a ring of radius $r_1 < r < r_2 = 0.92\arcsec$ where we computed a mean value. On the right we plot the difference of the initial ($0\,\degr$) and $90\,\degr$ rotation images where we cut a small window to assess a mean residual value. The ratio between the mean residual value and the mean value give us a level of symmetry. \textit{Bottom panels:} same process at $1.64\,\mu m$ with a ring of radius $r_1 < r < r_2 = 1.1\arcsec$. The scale intensity is logarithmic.}
\label{symmetry}
\end{figure}

This method is useful for studying the morphology of the CSE, in particular whether or not it has a central symmetry, like one expects for a uniform shell structure or a face-on disk; or an asymmetrical shape, like in an inclined disk-like structure for instance. Despite adaptive optics artefacts, the images in Fig.~\ref{env} seem to present a uniform intensity distribution in agreement with a shell or a face-on disk structure. More precisely, we estimated a symmetry level using a residual map obtained by subtracting 90 degree rotated versions of themselves from the images presented in Fig. 7. The result of this subtraction is shown in Fig.~\ref{symmetry}. In the initial image we assessed the average value in a ring, while in the subtracted image ($0\degr - 90\degr$) we compute an average value in a small window (see Fig.~\ref{symmetry}). Making the ratio of these two averages (i.e. the average value in the small window over the average value in the ring), we exclude a departure from symmetry larger than $7\,\%$ in NB\_1.64 and $6\,\%$ in IB\_2.18.

\section{Conclusion}

We present a study of the close environment (1000-2000 AU) of the Cepheid RS~Pup using near-infrared AO-assisted imaging. We used two different techniques to 1) measure the photometric emission due to the CSE and 2) obtain an image proportional to the CSE through a statistical analysis of the large number of short exposures in which speckles are still visible.

Using a simple image modeling method, based on the unresolved PSF calibrator, we estimated (Sect.~\ref{shiftandadd}) a likely photometric emission of 38\,$\pm$\,17\,\% and 24$\,\pm$\,11\,\% of the star flux in the NB\_1.64 and IB\_2.18 bands respectively for RS~Pup's CSE. Apart from the actual interest to better understand the mass loss in Cepheids, the presence of such an extended emission may have an impact on the distance estimate with the Baade-Wesselink method.

Using an original statistical study of our short exposure data cubes, we qualitatively showed (Sect.~\ref{morphological-analysis}) that the envelope in the two bands (NB\_1.64 and IB\_2.18) has a centro-symmetrical shape (excluding a departure from symmetry larger than $7\,\%$), suggesting that the CSE is either a face-on disk or a uniform shell.

Our observations were obtained in two filters isolating the Brackett 12--4  ($1.641\,\mu$m, NB\_1.64 filter) and Brackett $\gamma$ 7--4 ($2.166\,\mu$m, IB\_2.18 filter) recombination lines of hydrogen. We therefore propose that our observations show the presence of hydrogen in the CSE of RS~Pup. The star is naturally relatively faint in these spectral regions, because of deep hydrogen absorption lines in its spectrum. This appears as a natural explanation to the high relative contribution from the CSE compared to the stellar flux. Another contribution to the CSE flux may also come from free-free continuum emission, or Rayleigh scattering by circumstellar dust. However, this contribution is probably limited to a few percents of the stellar flux. The high relative brightness of the detected CSE therefore implies that the spectral energy distribution of the detected CSE is essentially made of emission in hydrogen spectral lines instead of in the form of a continuum. RS~Pup's infrared excess in the broad NB\_1.64 and IB\_2.18 bands is limited to a few percents \citep{Kervella-2009-05}, which excludes a continuum CSE emission of 40\% in the NB\_1.64 filter and 24\% in the IB\_2.18 filter as measured in our NACO images.

Our discovery is also remarkably consistent with the non-pulsating hydrogen absorption component detected in the visible H$\alpha$ line of RS~Pup by \citet{Nardetto-2008-10-1}, together with a phase-dependent component in emission. The extended emission we observe in the infrared is therefore probably caused by the same hydrogen envelope as the one that creates the H$\alpha$ emission observed in the visible. It should also be mentioned that our result is consistent with the proposal by \citet{Kervella-2009-05} that the envelope of RS~Pup is made of two distinct components: a compact gaseous envelope, which is probably at the origin of the observed NACO emission, and a very large ($\approx\,2\arcmin$) and cold ($\approx\,45$\,K) dusty envelope \citep{Kervella-2008-03} that is mostly transparent at near-infrared wavelengths owing to the reduced scattering efficiency. These two envelopes appear to be largely unrelated because of their very different spatial scales. However, it should be noted that \citet{Marengo-2009-01}  recently detected large envelopes around several Cepheids using Spitzer. The presence of hydrogen close to the Cepheids of their sample could be investigated at high angular resolution using the observing technique presented here.


 \begin{acknowledgements}
We thank Drs. Jean-Baptiste Le Bouquin and Guillaume Montagnier for helpful discussions. We also thank the ESO Paranal staff for their work in visitor mode at UT4. We received the support of PHASE, the high angular resolution partnership between ONERA, Observatoire de Paris, CNRS, and University Denis Diderot Paris 7. This work made use of the SIMBAD and VIZIER astrophysical database from CDS, Strasbourg, France and the bibliographic informations from the NASA Astrophysics Data System. Data processing for this work have been done using the Yorick language, which is freely available at http://yorick.sourceforge.net/.
\end{acknowledgements}


\bibliographystyle{aa}   
\bibliography{bibliographie}

\end{document}